%% file: solid-test-bench-lal.tex
\documentclass[a4paper,11pt]{article}
\pdfoutput=1 

\usepackage{jinstpub} 
\usepackage{siunitx}
\usepackage{tikz}
\usepackage{multirow}
\usepackage{ulem}

\newcommand{\up}[1]{\textsuperscript{#1}}
\newcommand{\down}[1]{\textsubscript{#1}}

\title{\boldmath Optimisation of the scintillation light collection and uniformity for the SoLid experiment}

\input{SoLid_author_list_june2018}

\emailAdd{bongrand@lal.in2p3.fr}

\abstract{
  This paper presents a comprehensive optimisation study to maximise the light collection efficiency of scintillating cube elements used in the SoLid detector. Very short baseline reactor experiments, like SoLid, look for active to sterile neutrino oscillation signatures in the anti-neutrino energy spectrum as a function of the distance to the core and energy. Performing a precise search requires high light yield of the scintillating elements and uniformity of the response in the detector volume. The SoLid experiment uses an innovative hybrid technology with two different scintillators: polyvinyltoluene scintillator cubes and \up{6}LiF:ZnS(Ag) screens. A precision test bench based on a \up{207}Bi calibration source has been developed to study improvements on the energy resolution and uniformity of the prompt scintillation signal of antineutrino interactions. A trigger system selecting the 1~MeV conversion electrons provides a Gaussian energy peak and allows for precise comparisons of the different detector configurations that were considered to improve the SoLid detector light collection. The light collection efficiency is influenced by the choice of wrapping material, the position of the \up{6}LiF:ZnS(Ag) screen, the type of fibre, the number of optical fibres and the type of mirror at the end of the fibre. This study shows that large gains in light collection efficiency are possible compared to the SoLid SM1 prototype. The light yield for the SoLid detector is expected to be at least 52$\pm$2 photo-avalanches per MeV per cube, with a relative non-uniformity of 6~\%, demonstrating that the required energy resolution of at least 14~\% at 1~MeV can be achieved.
}


\begin{document}
\maketitle
\flushbottom


\section{Introduction}
\label{sec:intro}

\begin{figure}[htbp]
  \centering
  \includegraphics[width=.45\textwidth]{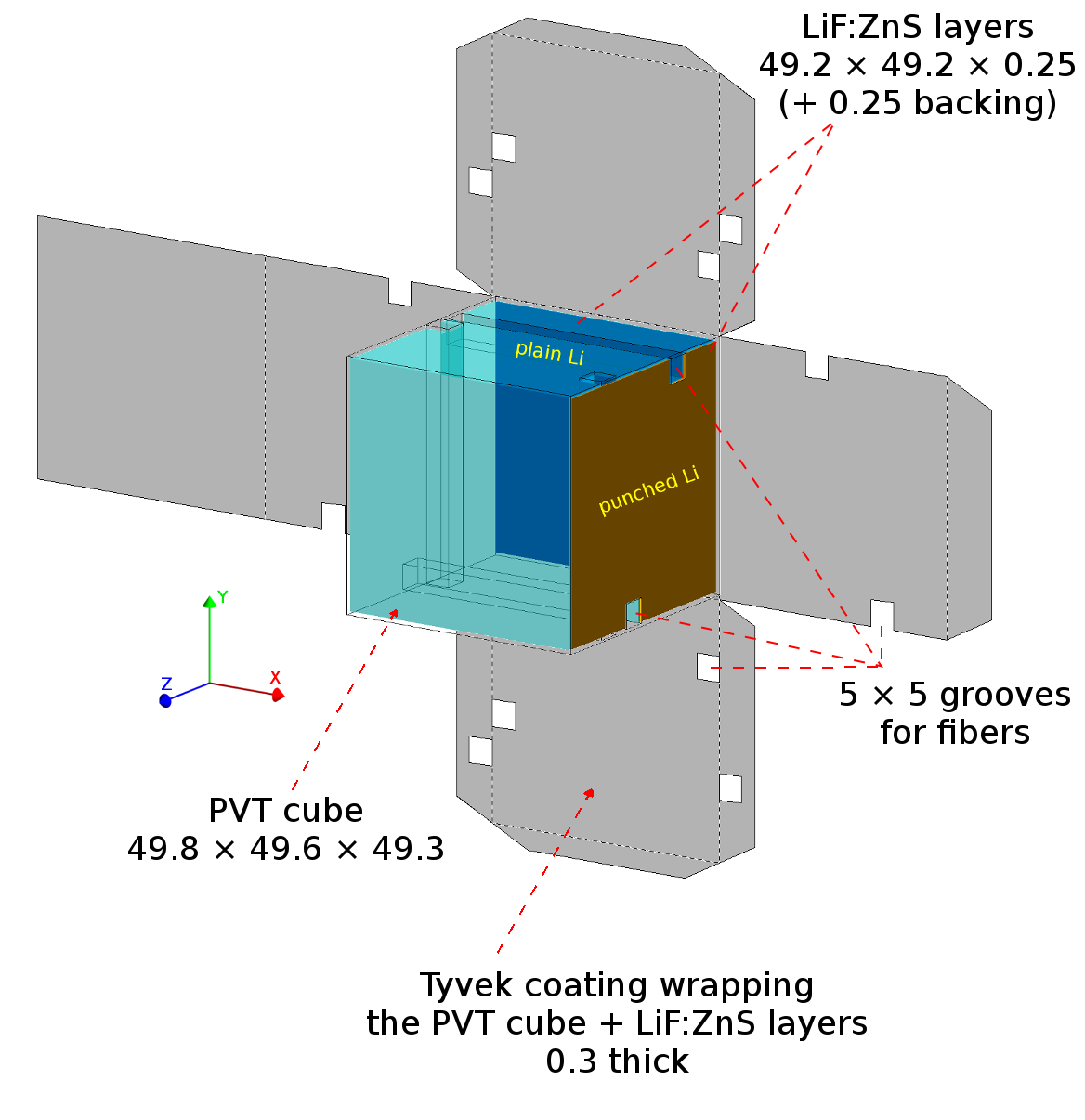}~~~
  \includegraphics[width=.45\textwidth]{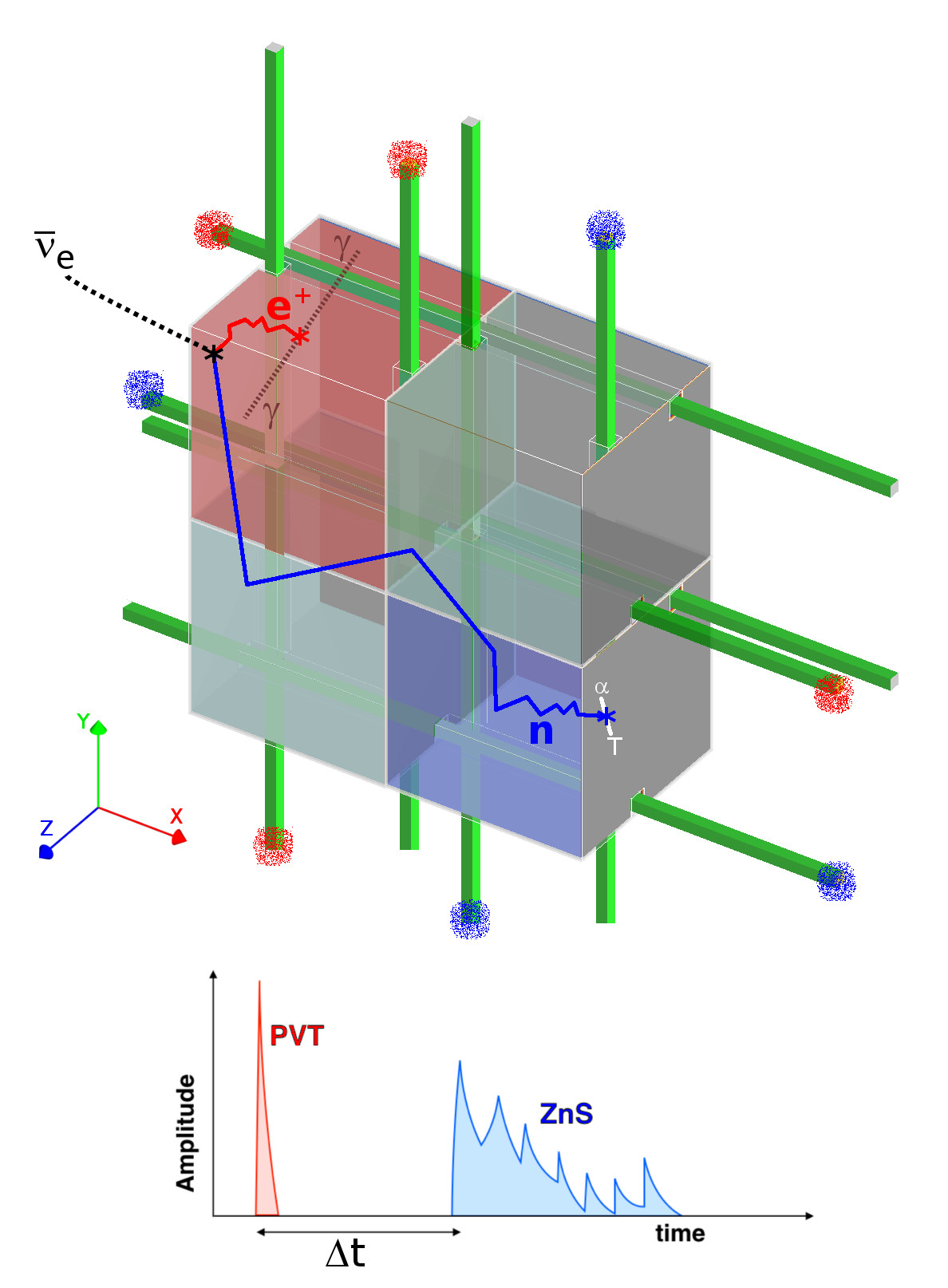}
  \caption{\label{fig:solid-principle} Principle of SoLid hybrid scintillators technology and antineutrino detection. (Left) The PVT cube is covered with two screens of \up{6}LiF:ZnS(Ag) and then wrapped in Tyvek reflective material. (Top right) Principle of $\overline{\nu}_e$ detection in a volume made of separated cubes: wavelength shifting fibres placed in perpendicular orientations are used to collect the scintillation light from each cube of the array. (Bottom right) Illustration of the time spectra of PVT and ZnS(Ag) scintillation used to identify the IBD reaction products.}
\end{figure}

SoLid very short baseline reactor antineutrino experiment \cite{solid} will search for active to sterile antineutrinos oscillations between 6 and 9~m of the BR2 research reactor at the SCK\raisebox{-0.8ex}{\scalebox{2.8}{$\cdot$}}CEN in Mol, Belgium. It consists of a novel fine segmented hybrid scintillator detector technology made of optically isolated polyvinyltoluene (PVT) plastic scintillator cubes, each coupled to neutron sensitive inorganic scintillator \up{6}LiF:ZnS(Ag) screens as illustrated in figure~\ref{fig:solid-principle}. These two different scintillators are used to detect both the positron and the neutron produced by the inverse beta decay (IBD) interaction of an antineutrino. The scintillation signals from the two scintillators are collected via the same wavelength shifting fibres connected with silicon Multi-Pixel Photon Counter from Hamamatsu (MPPCs\up{\texttrademark}). Interactions in each scintillator can easily be distinguished because of the different decay time structure of the two signals. The high level of segmentation given by the 5$\times$5$\times$5~\si{\cubic\centi\meter} detector elements provides an unprecedented granularity for reconstructing the antineutrino energy with a limited energy contamination of the 511~keV $\gamma$-rays coming from the IBD positron annihilation. Combined together, the robust neutron capture identification, signal localization and event reconstruction should allow the SoLid experiment to perform a precise very short baseline reactor based antineutrino oscillation search. This fine segmented plastic scintillator and optical fibre readout technology is also considered for other neutrino experiments, like the T2K near detector for example \cite{t2k}.

Plastic scintillator cubes in the SoLid experiment primarily serve as the antineutrino target since they contain a large number of free protons in the form of hydrogen nuclei. At the same time, it allows the measurement of the positron energy deposition, which in turn is related to the neutrino energy. The SoLid experiment uses ELJEN Technology EJ-200 PVT scintillator, which is one of the most efficient plastic scintillators with a light yield of around 10,000 photons per MeV. Light around 425 nm wavelength is produced with a decay time of 2.1 ns. Its refractive index is 1.58.

The SoLid neutron screen (NS) is a \up{6}LiF:ZnS(Ag) scintillator from Scintacor. The neutron capture on \up{6}Li produces two nuclei \up{3}H and \up{4}He sharing a kinetic energy of 4.78~MeV. This energy is converted into scintillation light which enters in the PVT cube and is subsequently collected by the optical fibres. The \up{6}LiF:ZnS(Ag) scintillator emits light with a maximum emission at 450 nm, close to the PVT emission, so the collection will be similar as for the plastic scintillator. It is a slower scintillator with a decay time of about 80~\si{\micro\second}. This time difference makes it easy to distinguish between light produced in the PVT and in the NS scintillators (figure~\ref{fig:solid-principle} bottom right). The NS has a thickness of about 250~\si{\micro\meter} and a molecular LiF to ZnS ratio of 1:2.

The assembly of the PVT and the \up{6}LiF:ZnS(Ag) scintillators is wrapped in Tyvek reflective material to optically isolate each scintillator cube of the detector in order to be able to locate the position of the IBD interaction. Additionally, the wrapping also acts as a reflector, increasing the light collected by the fibres. Squared holes and grooves of 5$\times$5~\si{\square\mm} section in the Tyvek and on the surface of the PVT scintillators allows for the optical fibres to go through the cubes and collect the scintillation light of both scintillators.

SoLid optical fibres are 3$\times$3~\si{\square\milli\meter} squared fibres of less than 1~m long produced by Saint-Gobain under the reference BCF-91A. The shape and dimensions of these fibres are well adapted to the Hamamatsu MPPCs S12572-050P 3$\times$3~\si{\square\milli\meter} light readout. A single MPPC is used per optical fibre. To increase the light-yield, a reflective aluminum mirror is thus used at the other end of the fibre. MPPCs and mirrors are connected to the fibres with optical grease (BC 630) and supported by 3D printed connectors. The fibres extremities are polished by the producer. The fibres have a polystyrene core, an acrylic cladding and a fluor-acrylic cladding in the case of double-clad fibres. The refractive indexes of these parts are respectively 1.60, 1.49 and 1.42. The BCF-91A optical fibres have been selected because they match both the PVT emission spectrum as well as the MPPC spectral response. These fibres shift blue light to green with absorption at 420~nm and emission peaking around 494~nm. The MPPC photon detection efficiency is maximal with 35~\% at 450 nm but it is almost the same at 500 nm. The decay time constant of the emitted light of 12~ns is much shorter than the time difference between positron and neutron signals in the SoLid detector. This time difference is dominated by the thermalisation and capture of the IBD neutron, which takes several tens of micro seconds \cite{solid, sm1}.

\begin{figure}[htbp]
  \centering
  \includegraphics[width=.4\textwidth]{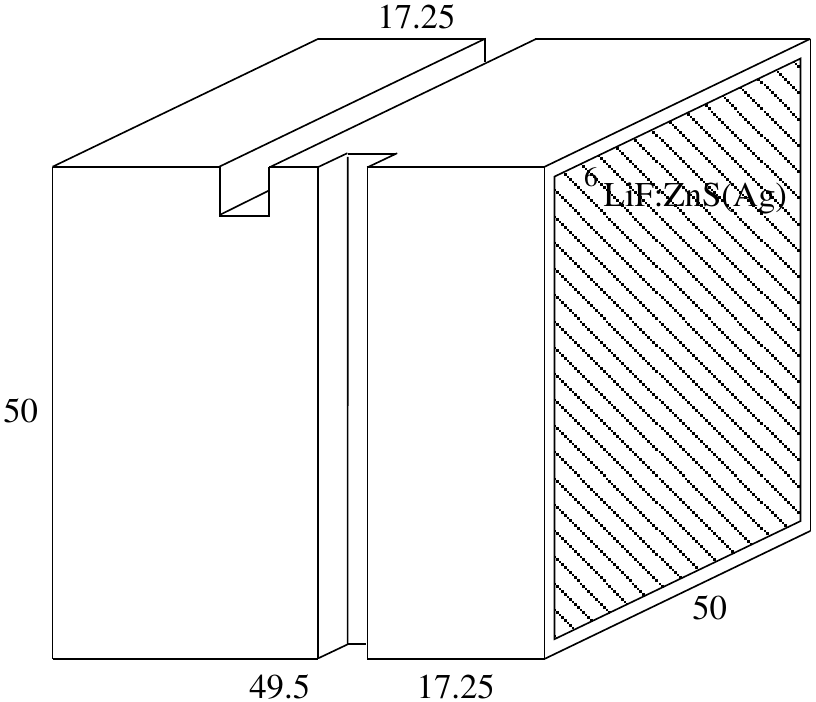}\hspace{.1\textwidth}
  \includegraphics[width=.4\textwidth]{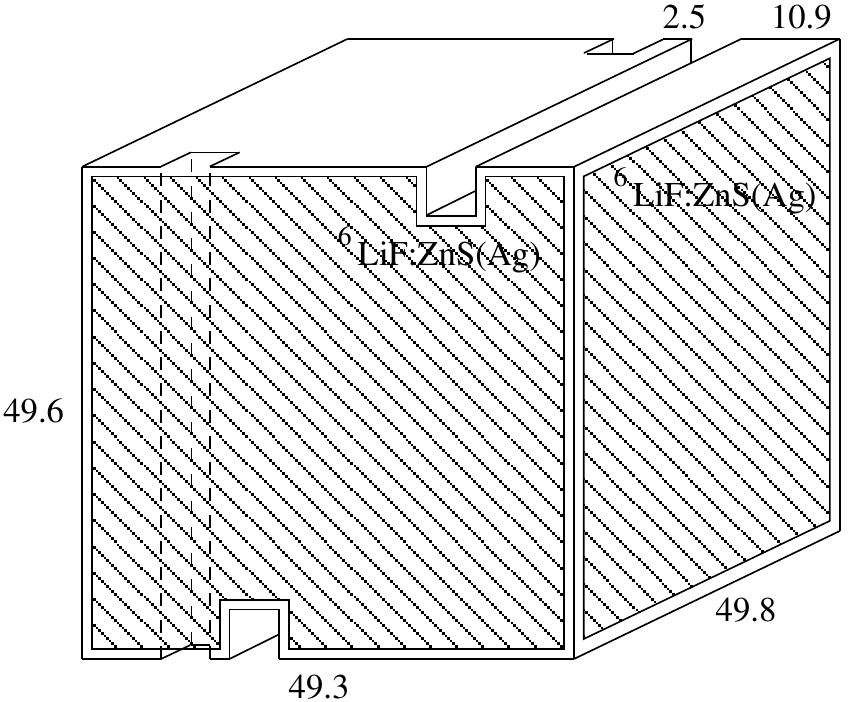}
  \caption{\label{fig:cube-designs} Design of the SM1 (left) and the SoLid Phase~1 (right) PVT scintillator cubes with two and four 5$\times$5~\si{\square\mm} fibre grooves respectively. The position of the \up{6}LiF:ZnS(Ag) screens is indicated. Lengths are given in mm.}
\end{figure}

A real scale prototype with an active mass of 288~kg, called SM1, was built and deployed at BR2 in 2014-2015 to demonstrate the antineutrino detection capabilities and background rejection \cite{sm1}. The module consisted of 9 planes of 16$\times$16 cubes. A cost-effective cube readout scheme was chosen with two single-clad optical fibres with one MPPC per fibre and a mirror at the other end of the fibre (figure~\ref{fig:cube-designs} left). A single screen of \up{6}LiF:ZnS(Ag) scintillator was installed per cube. First measurements of the light yield were performed resulting in 12 photo-avalanches (PA) per fibre. For the scintillator cube in this two fibres configuration, it would correspond to a stochastic energy resolution of 20~\%/$\sqrt{E({\mathrm{MeV}})}$. This first deployment validated the hybrid scintillator technology and the effect of the fine segmentation to discriminate and reduce the main experimental backgrounds.

The next phase of the SoLid experiment, called Phase~1, consists of a 1.6~t detector which has been constructed in 2017 and is now taking data. For a precise and timely antineutrino oscillation search, the aim of Phase~1 is to reach a stochastic term of the energy resolution $\sigma_E/E$ of at least 14~\% at 1~MeV. This would require to collect at least 50~PA/MeV/cube summing the light yield from all the fibres in a cube and after correcting for effects such as cross-talk in the MPPCs. For this purpose the number of optical fibre per cube has been doubled for this new detector (figure~\ref{fig:cube-designs} right). A second \up{6}LiF:ZnS(Ag) screen has also been added per cube in order to increase neutron detection efficiency and to reduce the capture time. This paper will present the studies and improvements in terms of light yield compared to the SM1 prototype in order to achieve these performances.

\section{Test bench setup}
\label{sec:test-bench}

The setup presented here has been inspired by the trigger system of an electron spectrometer \cite{spectrometer} used for the NEMO-3 and SuperNEMO experiments to qualify the plastic scintillators \cite{nemo3, supernemo-calo} and the regular deployment of \up{207}Bi sources in those detectors to produce the absolute energy calibrations. The principle of this setup is to use a \up{207}Bi calibration source and a trigger system to produce mono-energetic conversion electrons (see section~\ref{subsec:source}) in order to compare different detector element configurations. The setup is also capable of giving the absolute light yield to determine the energy scale and energy resolution of the PVT detection elements. It has been designed to be as flexible as possible in order to test various configurations for the SoLid scintillator cubes: wrapping, position and type of fibres, effect of the \up{6}LiF:ZnS(Ag) screen, machining and cleaning of the cubes, MPPCs and fibre reflectors. The test bench has been installed in a polyethylene black box (120$\times$120$\times$20~\si{\cubic\centi\meter}) sufficiently large to accommodate the full length of the SoLid fibres in both X and Y directions. The setup is installed in an air-conditioned room at a temperature of around 19~\si{\degreeCelsius}.

Since the \up{207}Bi radioactive source is mainly emitting $\gamma$ particles, it is necessary to use a triggering system to select only the conversion electrons entering the cubes (see section~\ref{subsec:source}). Otherwise, the signal will be dominated by Compton-scattering of $\gamma$-rays and the energy spectrum would give a lower precision on the light yield measurements than the peak from conversion electrons. The triggering system is described in section~\ref{subsec:trigger}. The energy spectrum and losses in the materials have been studied with Monte-Carlo simulations and are presented in section~\ref{subsec:simulation}.

\begin{figure}[htbp]
  \centering
  \includegraphics[width=.85\textwidth]{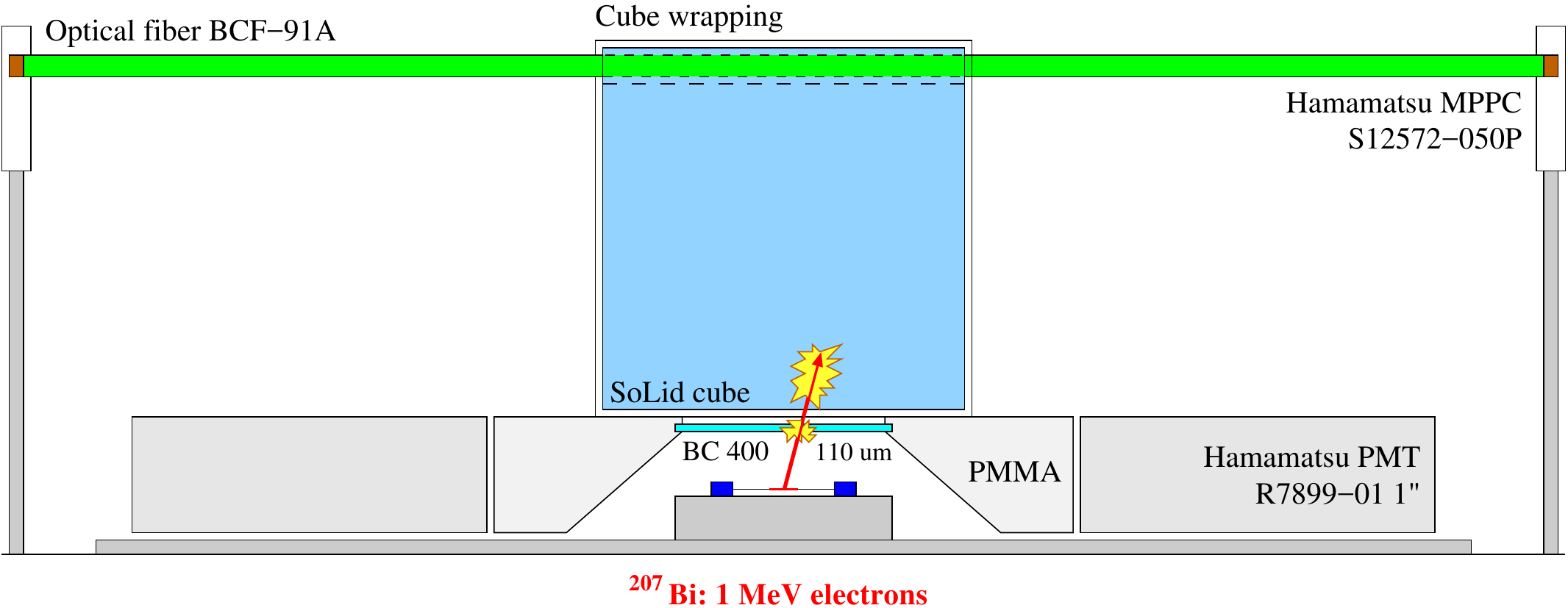}
  \caption{\label{fig:test-setup} Schematic description of the scintillator test setup in the standard configuration used for most of the measurements (a single wrapped cube along one optical fibre with double end MPPC readout). The calibration source, the PMTs and the scintillator cube are mounted on a rail in order to allow moving the system along the fibre.}
\end{figure}

In order to make comparisons between the different measurements and to simplify the operations a standard configuration has been defined for the main tests. This configuration is presented in figure~\ref{fig:test-setup}. It consists of a single SoLid scintillator cube (almost always the same for this publication) with its Tyvek wrapping of thickness 270~\si{\micro\meter} and read out by a single fibre and one MPPC at each end. The MPPCs are supplied with an over-voltage of 1.5~V, which is the bias voltage applied to operate the MPPC. This setting balances gain and cross-talk for this generation of photo-detectors (see section~\ref{subsec:xtalk}). The uncertainties on the measurements are discussed in section~\ref{subsec:uncertainties}.

The scintillator cubes and the triggering system are both mounted on a rail and can be moved with a light tight manual jack from outside the black box. This design allows for moving the full system along the fibre to be able to measure the light attenuation for different cube positions along the fibre (see section~\ref{subsec:optical-fibres}). In the case of the SM1 prototype the thickness of the Tyvek wrapping allowed for scintillation light to pass through the wrapping. However, the wrapping of neighbouring cubes allowed to recover a fraction of the light otherwise lost to the neighbouring environment. The rail allows then to perform measurements with a series of 16 cubes connected to a single fibre, which is closer to a realistic detector configuration (see section~\ref{sec:comparison-sm1-solid}).

\subsection{The \up{207}Bi radioactive source}
\label{subsec:source}

The \up{207}Bi isotope is well suited to test the SoLid scintillator performance in term of the energy scale and resolution since it produces mono-energetic electrons around 1~MeV. This is the same order of magnitude as the antineutrino energy determined from the positron energy deposit, which is between the IBD threshold of 1.806~MeV and 8~MeV. As already mentioned the detected 1~MeV Gaussian peak allows accurate comparisons between different detector configurations.

\begin{figure}[htbp]
  \centering
  \begin{tikzpicture}
    \node at (2,0.5) {$^{207}_{~82}$Pb};
    \draw [black,ultra thick] (0,1) -- (4,1); 
    \node [left] at (4.1,1.2) {\small 0};

    \draw [->,black,thick] (3,1.95) -- (3,1.03);
    \draw [black,thick] (0,1.95) -- (4,1.95);
    \node [left] at (4.1,2.15) {\small 570};
    \node [rotate=55,above] at (3.3,2.35) {\footnotesize 99.9~\%};
    \node [below,blue] at (3.35,1.95) {\footnotesize 570};

    \draw [->,black,thick] (2,3.72) -- (2,1.98);
    \draw [black,thick] (0,3.72) -- (4,3.72);
    \node [left] at (4.1,3.92) {\small 1633};
    \node [rotate=55,above] at (2.3,4.07) {\footnotesize 84.1~\%};
    \node [below,blue] at (2.4,3.72) {\footnotesize 1064};

    \draw [->,black,thick] (1,4.9) -- (1,1.98);
    \draw [black,thick] (0,4.9) -- (4,4.9);
    \node [left] at (4.1,5.1) {\small 2340};
    \node [rotate=55,above] at (1.3,5.25) {\footnotesize 6.90~\%};
    \node [below,blue] at (1.4,4.9) {\footnotesize 1770};

    \draw [black,ultra thick] (5,5) -- (6.5,5); 
    \node at (5.75,5.5) {$^{207}_{~83}$Bi};
    \node [right] at (6.5,5) {\small $Q_\epsilon$ : 2398};
    \node at (5.75,4.7) {\small $\epsilon$ : 99.93~\%};
    \node at (5.75,4.2) {\small T$_{1/2}$ : 32.9 y};
  \end{tikzpicture}
  \caption{\label{fig:207bi} \up{207}Bi decay scheme to excited states of \up{207}Pb with energies given in~keV \cite{207bi}.}
\end{figure}
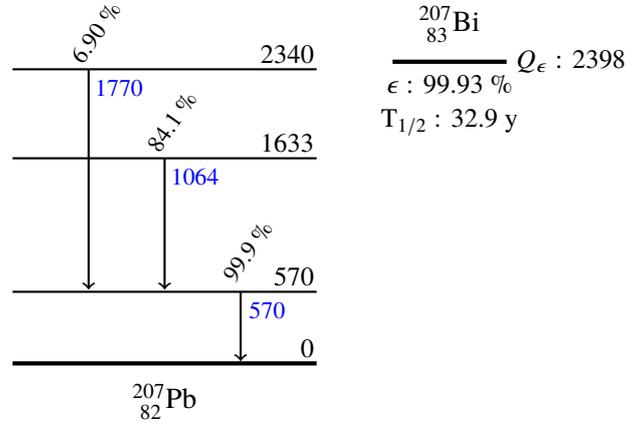

The \up{207}Bi isotope decays through electron capture almost exclusively to excited states of \up{207}Pb \cite{207bi}. The \up{207}Pb de-excitations occur through 3 main $\gamma$-ray emissions (570, 1064 and 1770~keV) as illustrated in figure~\ref{fig:207bi}. These $\gamma$-ray emissions could be replaced by atomic $K$, $L$ or $M$ shell conversion electrons as presented in table~\ref{tab:207bi}. The conversion electrons associated to the 1770~keV de-excitation are negligible and those associated to the 570~keV occur only in 1.5~\% of the decays over an important $\gamma$ background. Most of the useful conversion electrons are associated to the 1064~keV de-excitation and have an energy between 976 and 1060~keV with a total probability of 9.5~\%. Given the finite energy resolution of the SoLid detector (14-20~\%), only one main peak at an average energy of 995~keV is expected (see section~\ref{subsec:simulation}).

\begin{table}[htbp]
\centering
\caption{\label{tab:207bi} Main conversion electrons emitted by the de-excitation of \up{207}Pb after \up{207}Bi decays \cite{207bi}.}
\smallskip
\begin{tabular}{|c|c|c|c|}
  \hline
  \textbf{Transition} & \textbf{Shell} & \textbf{Energy (keV)} & \textbf{Probability (\%)} \\
  \hline
  \multirow{3}{*}{570 $\rightarrow$ 0} & $K$ & 482 & 1.5 \\
   & $L$ & 555 & 0.43 \\
   & $M$ & 566 & 0.12\\
  \hline
  \multirow{4}{*}{1633 $\rightarrow$ 570} & $K$ & 976 & 7.1\\
   & $L$ & 1049 & 1.8\\
   & $M$ & 1060 & 0.44\\
   & $N$ & 1063 & 0.12\\
  \hline
\end{tabular}
\end{table}

The \up{207}Bi source used in the setup has an activity of 37 kBq. The active material has been deposited between 2 mylar foils of 0.9~\si{\milli\gram\per\square\centi\meter}. The energy losses in these mylar foils is negligible compared to our detector energy resolution. The active area of the radioactive source represents a 5 mm diameter disk, which is small compared to the 5$\times$5~\si{\square\centi\meter} surface of the scintillator cube.

\subsection{The external triggering system}
\label{subsec:trigger}

The principle of the triggering system is to select only the 1~MeV mono-energetic conversion electrons by detecting them in the thin (110~\si{\micro\meter}) plastic scintillator (BC 400 - 2$\times$1~\si{\square\centi\meter}) before they enter the SoLid scintillator cube. The light produced in the trigger scintillator is collected by two polymethyl methacrylate (PMMA) light guides which channel the light towards two 1" PMTs (Hamamatsu R7899-01) equipped with custom made dividers developed for PMT tests at LAL Orsay. Good optical coupling is ensured by optical grease (BC 630) between the thin scintillator and the light-guides and by an optical epoxy silicone rubber compound (RTV 615) between the light-guides and the PMTs. The light collection of this setup is not sufficient to reconstruct precisely the energy deposited by the crossing electrons but detailed \textsc{Geant4} based simulations, described in section~\ref{subsec:simulation}, show that it represents negligible energy loss. This thin scintillator provides a triggering signal to tag the charged particle entering the cube. The triggering system has been designed to minimize the distance between the source and the scintillator cube in order to maximize the solid angle and reduce the energy loss of the electrons before they enter the cube.

\begin{figure}[htbp]
  \centering
  \includegraphics[width=.55\textwidth]{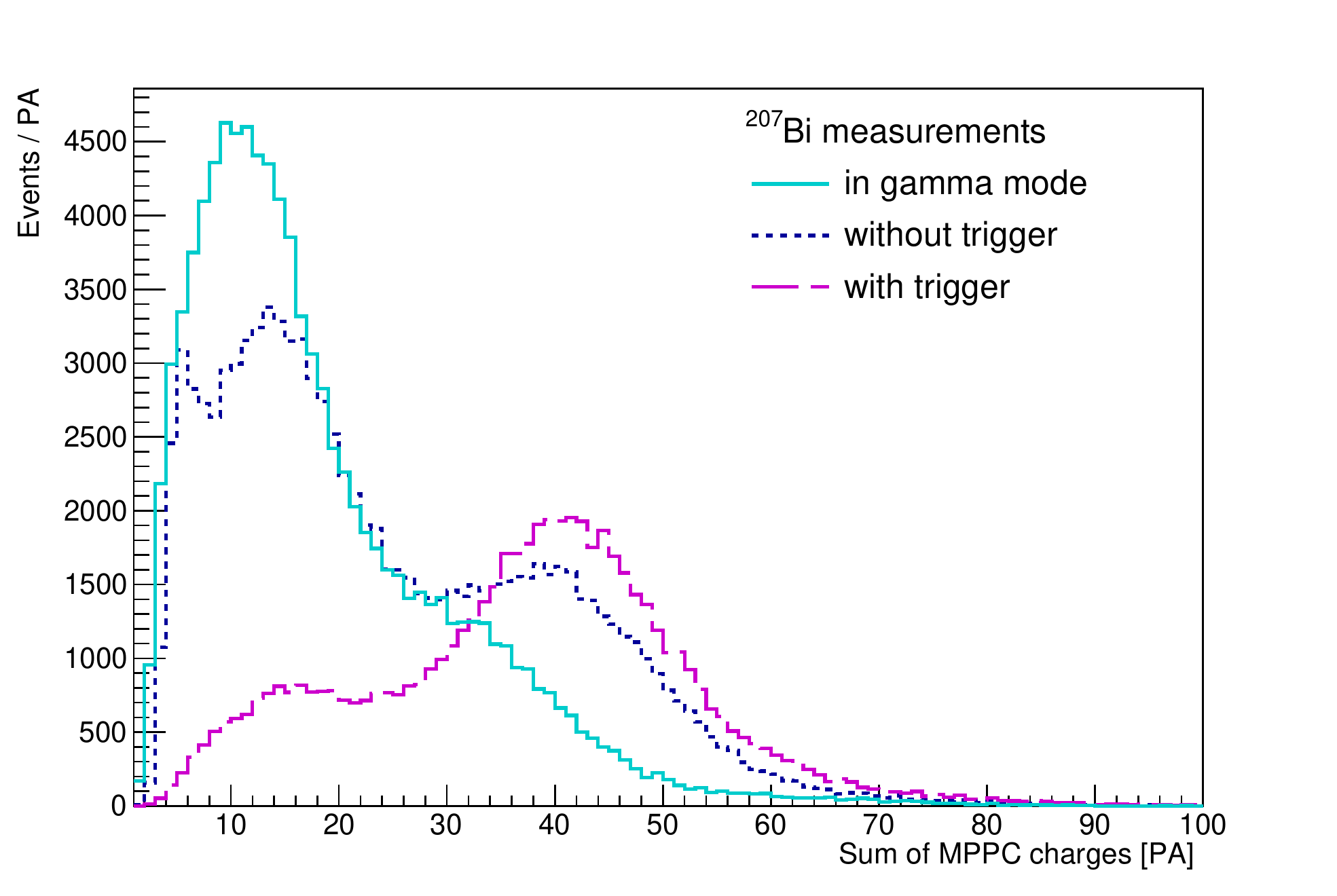}
  \caption{\label{fig:bi207-trigger-modes} Comparison of the \up{207}Bi energy spectra registered from the PVT cube in different triggering modes. In blue is the spectrum in coincidence with the two MPPCs only, in magenta the spectrum in coincidence with the 110~\si{\micro\meter} triggering scintillator and in cyan using this scintillator as an electron veto.}
\end{figure}

Figure~\ref{fig:bi207-trigger-modes} illustrates the impact of the triggering system for selecting the 1~MeV conversion electrons. The three spectra represented are obtained when triggering in coincidence on the 2 MPPCs only (gammas + electrons in blue), when triggering in coincidence with the small scintillator (electrons in magenta), and when using the small scintillator as an electron veto to select only the gammas (in cyan). The reconstruction of the energy deposited in the cube is explained in section~\ref{subsec:reconstruction}. The shape of the energy spectrum of the gammas is less sensitive to light collection improvement tests but still gives valuable information on the detector response to antineutrino interactions. Indeed the gammas are interacting in the whole volume of the scintillator while the conversion electrons will only interact in a small portion of the scintillator ($<$1~\si{\cubic\centi\meter}) in front of the source. The detector response to gammas is closer to the prompt signal from antineutrino interactions that will also occur in the whole volume of the scintillator. Less than 2~\% difference between the energy scale determination from the Compton edge fit and the 1~MeV peak is observed\footnote{The energy of the Compton edge for the 1064~keV $\gamma$ is 858~keV and because of energy losses and resolution the ``1~MeV'' peak is expected at 910~keV. The respective fitted values were 39.0 and 40.8 photo-avalanches.}. This is within the systematics uncertainties (section~\ref{subsec:uncertainties}) and shows that the average response through the scintillator volume is the same as the centre of the cube surface.

\subsection{Electronics and acquisition}
\label{subsec:electronics}

The photo-detectors selected for SoLid are the Hamamatsu MPPCs S12572-050P 3$\times$3~\si{\square\milli\meter}. These devices were not specifically studied in our setup. The measurements performed only concerned the cross-talk probability of the MPPCs (section~\ref{subsec:xtalk}) that needs to be accounted for light yield determination. The MPPCs were soldered on custom made PCBs installed in 3D printed supports also used to hold the optical fibre, as in the SM1 prototype. The optical contact between the MPPC and the fibre is made by optical grease (BC 630).

To supply voltage, amplify, shape and extract the MPPC signals, a custom made three channels prototype board is used. This board has been developed to validate the analog electronic boards of the SM1 prototype. The voltage is provided by two external power supplies (EA-PSI 6150-01): one at 65~V for the MPPC supply and one at 5~V for the amplifiers. These power supplies have a very good resolution of 10~mV and a stability of better than 5~mV. With this setup the same voltage is provided to all the channels. The two MPPCs have been selected to have close operating voltages ($V_{OP}$ = 67.40 and 67.46 V respectively). The two trigger PMTs are powered by an Ortec 556 power supply at -1400~V.

An eight channel waveform digitizer developed at LAL based on the WaveCatcher ASIC is capturing the signals from all photon detectors (\cite{wavecatcher, wavecatcher2}). This module is directly controlled by USB and a CVI software allowing to define the acquisition settings, perform analysis and store the digitized pulses. The trigger is set as a coincidence of the two negative PMT signals at -5 mV and the positive MPPC signals at 2 mV. The sampling is made over 1024 points at 1.6 GS/s to properly sample the waveforms over their whole pulse length. This corresponds to a 640~ns time window. More details on the reconstruction of the MPPC pulses and the energy are presented in section~\ref{sec:measurements}.

\subsection{Simulation of the setup}
\label{subsec:simulation}

Simulation studies were performed to determine the mean energy of the $\sim$1~MeV conversion electron peak from the \up{207}Bi source and to compute the energy losses in the thin triggering scintillator and the wrapping around the cubes. These simulations are using the Bayeux suite~\cite{bayeux} developed for the simulation of the SuperNEMO experiment, in conjunction with \textsc{Geant4}~\cite{geant4}.

\begin{figure}[htbp]
  \centering
  \includegraphics[width=.85\textwidth]{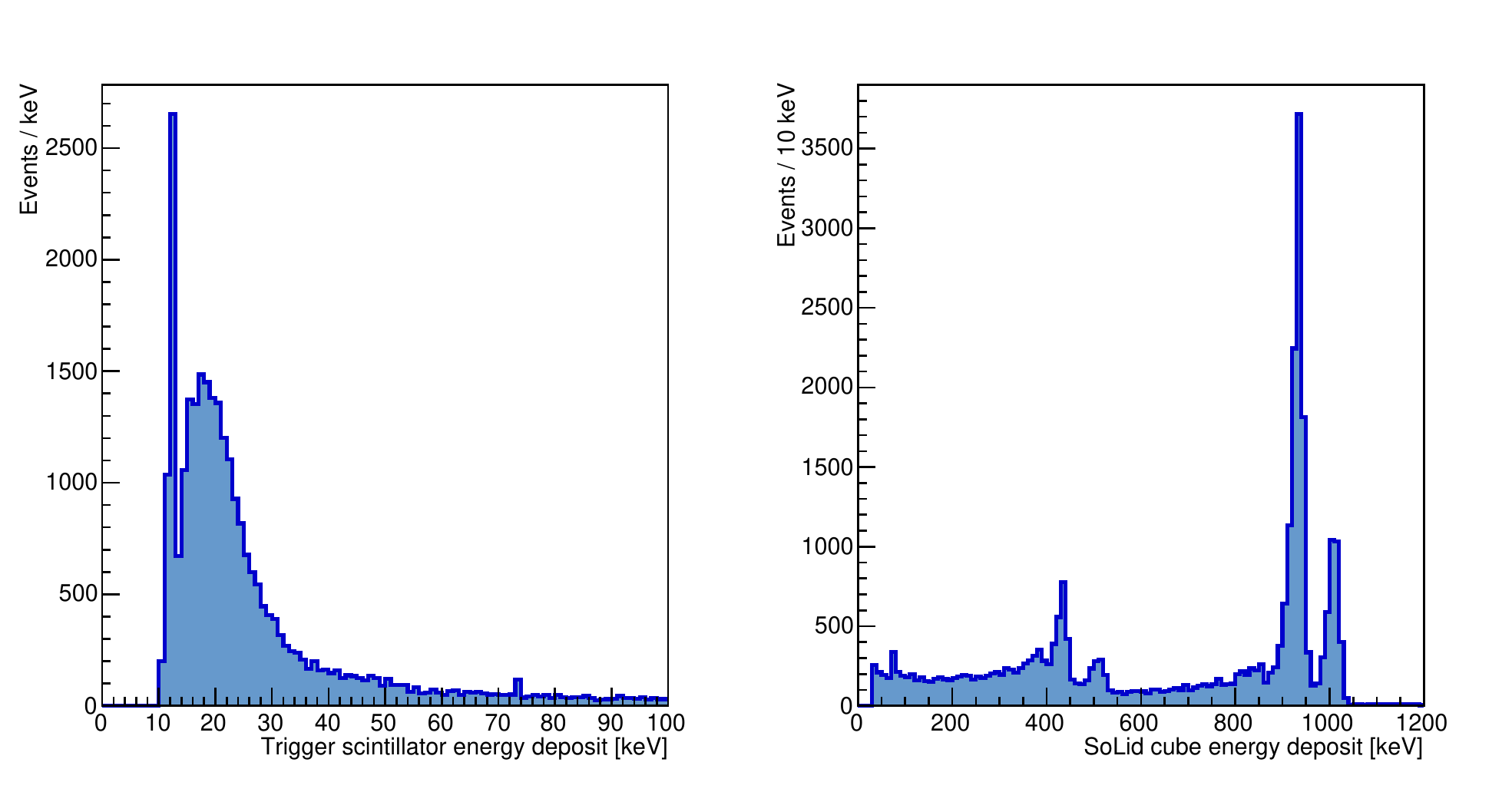}
  \caption{\label{fig:edeposit} Simulated perfect energy deposition by the \up{207}Bi conversion electrons in the 110~\si{\micro\meter} triggering scintillator (left) and in the SoLid PVT cube (right) wrapped in thick Tyvek (270~\si{\micro\meter}). The conversion electrons from the 570 and 1064~keV transitions (doubled by the K and L atomic shells) are visible on the right plot. The detector energy resolution is not applied here in the simulation.}
\end{figure}

\begin{figure}[htbp]
  \centering
  \includegraphics[width=.55\textwidth]{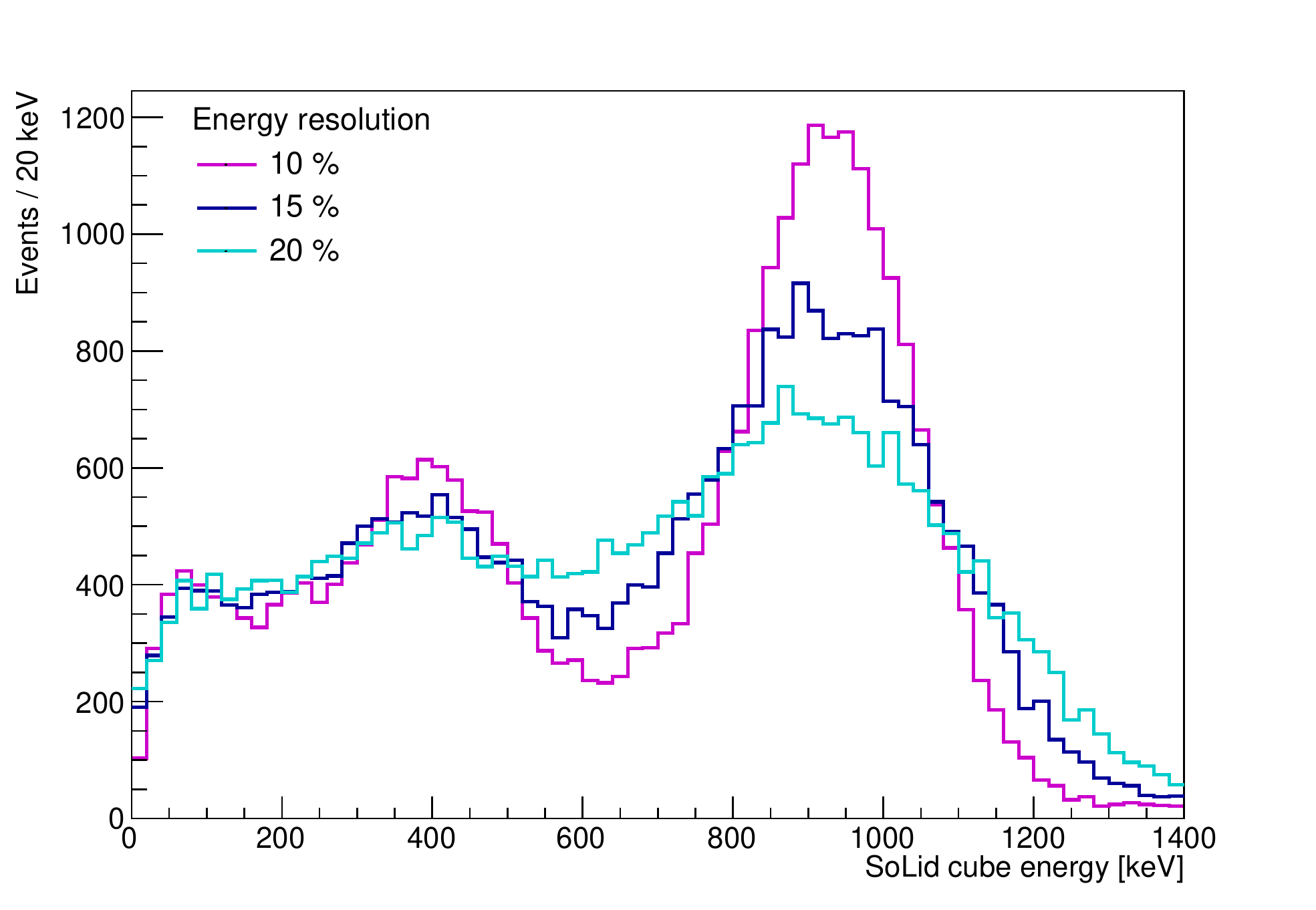}
  \caption{\label{fig:eresolution} Simulated energy deposition by the \up{207}Bi conversion electrons in the SoLid cube wrapped in 270~\si{\micro\meter} thick Tyvek after applying an energy resolution value of $\sigma_E/E =$ 10~\% (magenta), 15~\% (blue), 20~\% (cyan).}
\end{figure}

The result of this simulation indicates that on average only $\sim$25~keV is lost by the electrons in the triggering scintillator as can be seen in figure~\ref{fig:edeposit}. This is negligible compared to the conversion electron energy in the main peak (figure~\ref{fig:edeposit} right) and the energy resolution of the SoLid scintillator cubes. Applying the detector energy resolution to the simulation, one can see in figure~\ref{fig:eresolution} that the double-peak structure around 1 MeV disappears. Also the conversion peak around 500~keV is no longer visible over the Compton background of the 1064~keV $\gamma$-rays. It is therefore not possible to observe both conversion electron energy peaks distinctly. For this reason, the light yields will be determined by fitting the electron energy peak around 1~MeV by a gaussian function.

Different cube wrappings have been tested to improve the light reflectivity in the SoLid cubes. Tyvek\up{\textregistered} is the most suitable material to wrap the scintillator cubes for the SoLid experiment, as will be explained in section~\ref{subsec:wrapping}. In the simulation of this setup the Tyvek has been added as a uniform material of a given thickness and density around the cubes. This is an approximation since Tyvek, consisting of HDPE fibres, is non-uniform in thickness. In table~\ref{tab:tyvek} the properties of the Tyvek sheets used for the SM1 and SoLid Phase~1 detectors are presented. The respective average thicknesses are 205 and 270~\si{\micro\meter}. The ranges are estimates given by the producer DuPont\up{\texttrademark} based on the measurement of individual specimens. These values have been used to simulate different samples for estimating the electron energy loss before entering the cubes and to obtain the reference peak position to be compared to the measured values.

\begin{table}[htbp]
\centering
\caption{\label{tab:tyvek} Properties of the Tyvek\up{\textregistered} sheets from DuPont\up{\texttrademark} datasheets. We computed the densities from the average thickness and weight.}
\smallskip
\begin{tabular}{|c|c|c|c|c|}
  \hline
   \multirow{2}{*}{\textbf{Sample}} & \multirow{2}{*}{\textbf{Reference}} & \textbf{Weight} (\si{\gram\per\square\meter}) & \textbf{Thickness} (\si{\micro\meter}) & \textbf{Density} (\si{\gram\per\cubic\centi\meter})\\
   & & average [range] & average [range] & average\\
  \hline
  SM1 Tyvek\up{\textregistered} & DuPont\up{\texttrademark} 1073D & 75 [72 - 78] & 205 [135 - 275] & 0.366\\
  SoLid Tyvek\up{\textregistered} & DuPont\up{\texttrademark} 1082D & 105 [101.5-108.5] & 270 [190 - 350] & 0.389\\
  \hline
\end{tabular}
\end{table}

\begin{figure}[htbp]
  \centering
  \includegraphics[width=.49\textwidth]{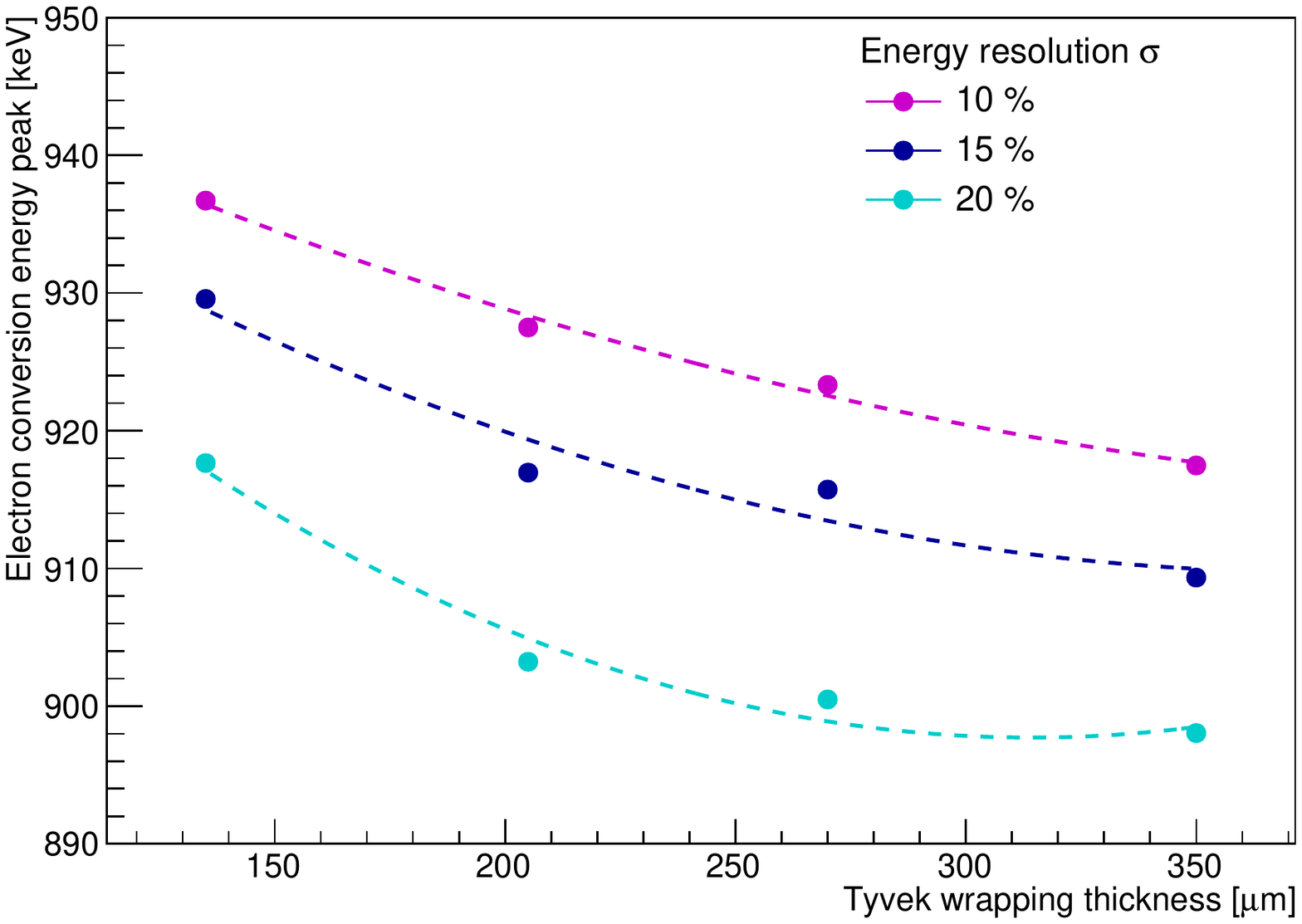}
  \includegraphics[width=.49\textwidth]{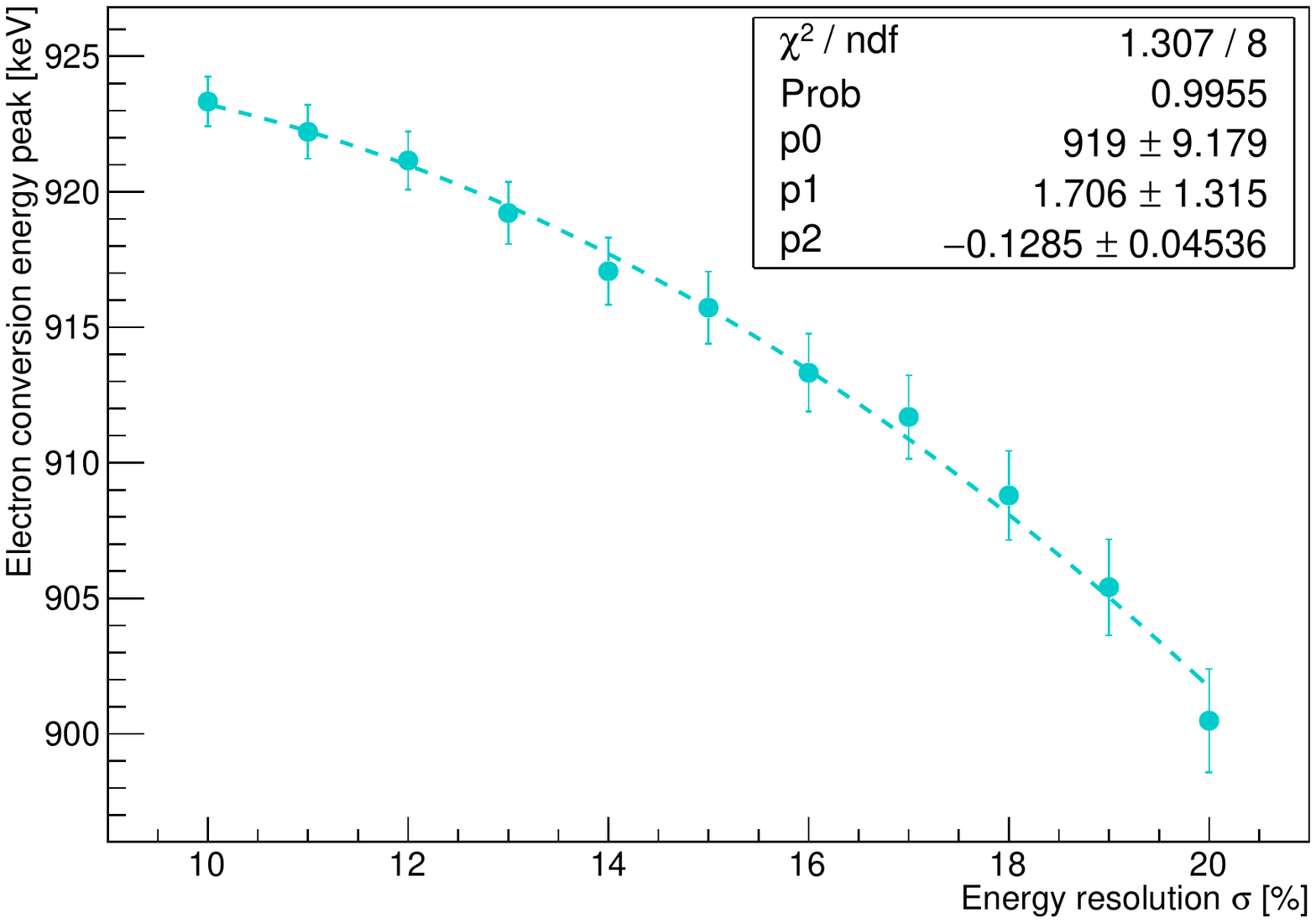}
  \caption{\label{fig:tyvek-thickness} Left: Fitted value of the simulated \up{207}Bi energy peak of SoLid cube wrapped in Tyvek as a function of Tyvek thickness after applying different energy resolution corrections ($\sigma_E/E = 10, 15, 20$~\%). Right: Variation of the fitted peak value as a function of energy resolution for 270~\si{\micro\meter} Tyvek. This graph is fitted by a second order polynomial function.}
\end{figure}

Figure~\ref{fig:tyvek-thickness} left shows that the energy loss in the Tyvek wrapping is also of the order of a few tens of~keV. The non-uniformity of the Tyvek wrapping should not influence the measurements since the average energy peak position is changing only $\sim$2~\% over the whole thickness range simulated. The difference in the fitted peak value as a function of energy resolution is due to the averaging over a different fraction of lower energy events seen before the electron conversion peak. For SM1 cube wrapping and a 20~\% energy resolution a calibration peak around 900~keV is obtained. For the SoLid Phase~1 cube wrapping and a 15~\% energy resolution a calibration peak of around 910~keV is obtained. The function fitted in figure~\ref{fig:tyvek-thickness} right, for the 270~\si{\micro\meter} SoLid Phase~1 Tyvek wrapping, will be used at each measurement to determine the energy peak position and the energy light yield in PA/MeV. The input energy resolution is first determined by the number of PA measured in the peak, as presented in section~\ref{subsec:analysis}.


\section{Measurements and data processing}
\label{sec:measurements}

\subsection{Pulse reconstruction}
\label{subsec:reconstruction}

\begin{figure}[htbp]
  \centering
  \includegraphics[width=.49\textwidth]{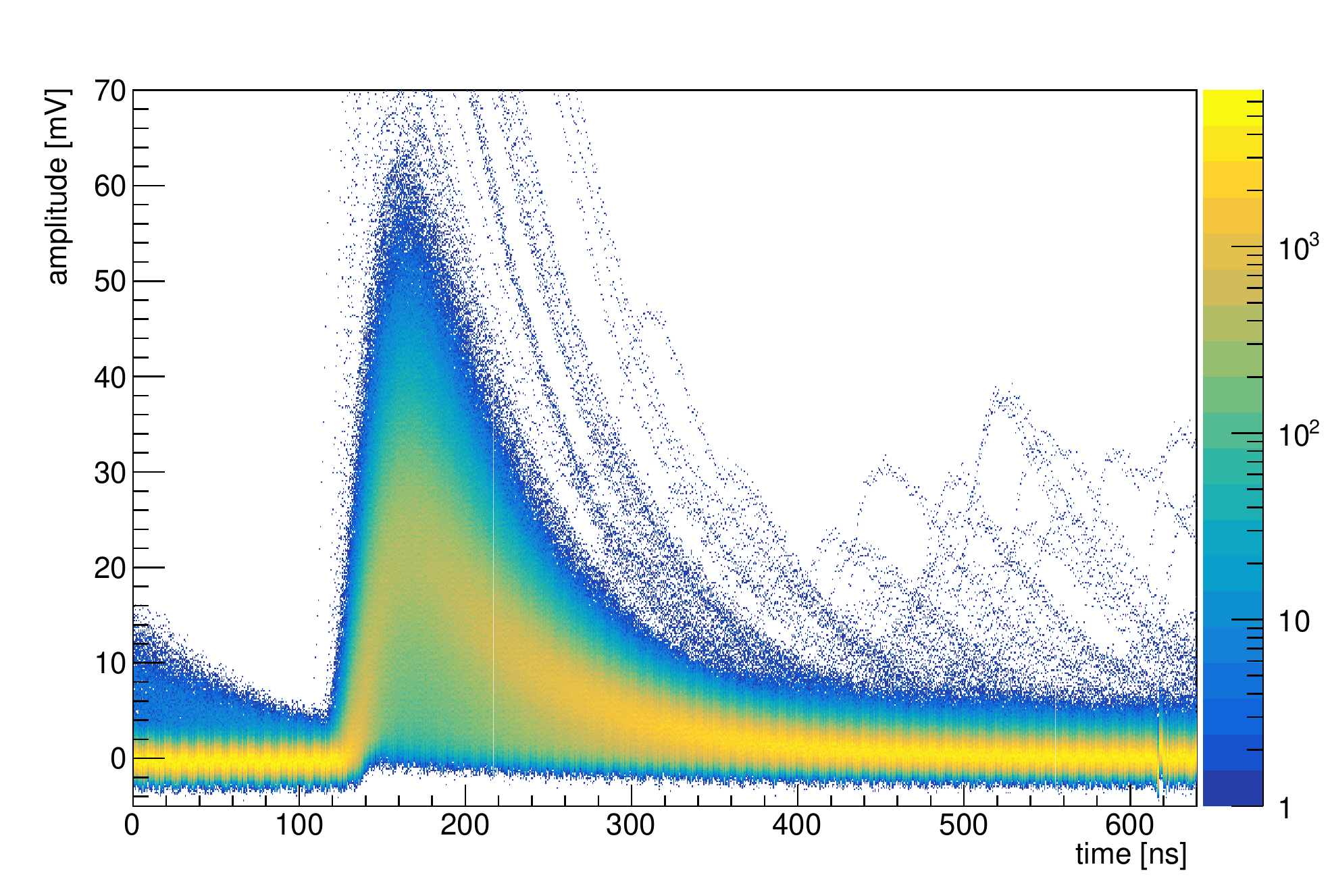}
  \includegraphics[width=.49\textwidth]{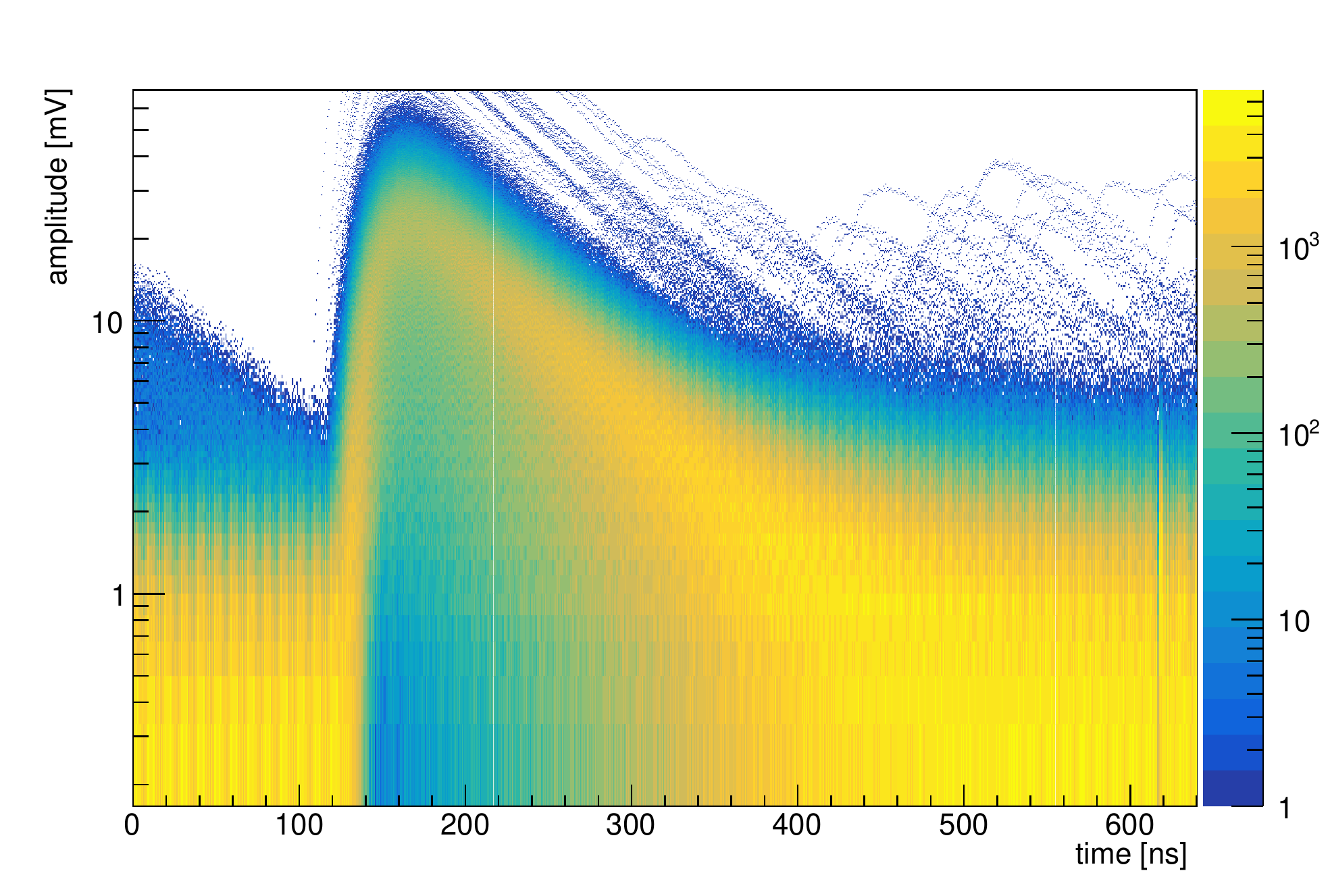}
  \caption{\label{fig:bi207-pulses} Persistence view of MPPC pulses registered from a SoLid scintillator cube during a \up{207}Bi measurement in linear (left) and log (right) scales. The few delayed pulses could be due to random coincidences or after-pulses.}
\end{figure}

The MPPC pulses reconstruction is done off-line from the 640~ns samplings registered by the acquisition. Figure~\ref{fig:bi207-pulses} shows a cumulated view of all the pulses registered during one \up{207}Bi run. The pedestal is computed before the rise of the pulse and the pulse integral is computed around the maximum amplitude. These parameters are both expressed in V~ns in the following. Pulse integral has a better resolution to individual PAs than the amplitude.

\subsection{MPPC cross-talk correction}
\label{subsec:xtalk}

Optical cross-talk occurs in MPPCs when during the primary avalanche multiplication some photons are emitted and start secondary avalanches in one or more neighbouring cells. Since a few tens of photons are emitted by a single avalanche, the cross-talk probability is high when no optical barrier (metallic trench) is implemented. This is the case for the generation of MPPCs used in the SoLid experiment, resulting in cross-talk probability of 10 to 30~\% depending on the over-voltage.

\begin{figure}[htbp]
  \centering
  \includegraphics[width=.55\textwidth]{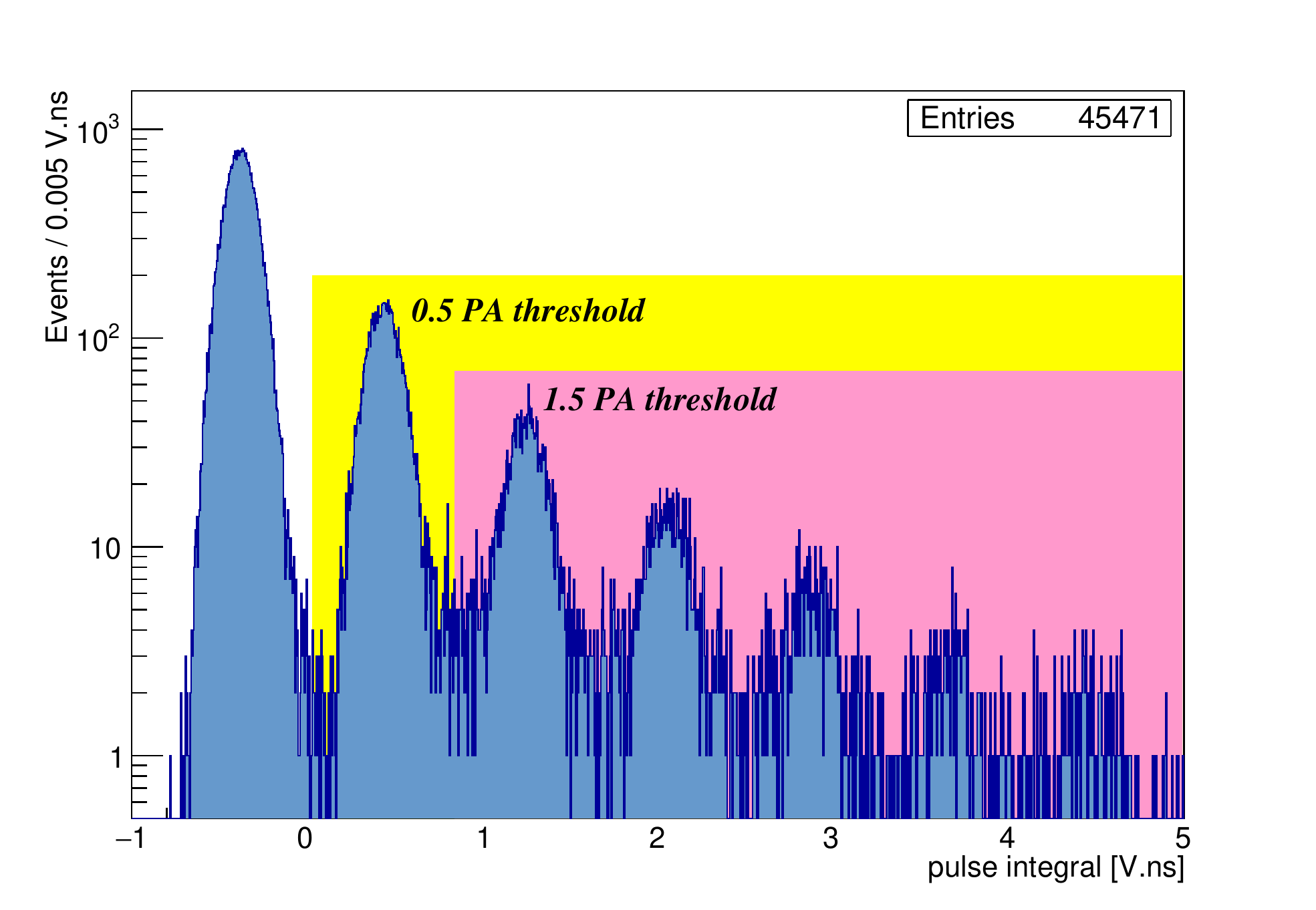}
  \caption{\label{fig:xtalk-calcul} Determination of the optical cross-talk probability of an MPPC from the dark count pulses. The first peak from left corresponds to the pedestal and the following peaks correspond to 1, 2, 3 or more PAs.}
\end{figure}

The optical cross-talk can be measured using dark count rate (DCR) pulses when the MPPCs are not connected to the fibre. Acquiring random trigger events only 1 PA signal peak should be observed from DCR. However, because of the optical cross-talk also peaks higher than the 1~PA peak are observed as shown in figure~\ref{fig:xtalk-calcul}. The cross-talk probability is measured as the ratio of the number of DCR events above the 1.5 and 0.5~PA thresholds, noted $N_{1.5 PA}$ and $N_{0.5 PA}$, as explained by the following equation:
\begin{equation}
  \label{eq:prob-xtalk}
  P_{cross-talk} = \frac{N_{1.5 PA}}{N_{0.5 PA}}
\end{equation}

These numbers of DCR events above each threshold are obtained by integrating the number of events in the peaks as illustrated on the figure~\ref{fig:xtalk-calcul} by the two coloured regions. At 1.5~V over-voltage we find on average 17.7~$\pm$~1.0 (stat)~\% for the two MPPCs.

\subsection{Procedure to calculate the light yield}
\label{subsec:analysis}

After the reconstruction of the pulses parameters several steps are still needed to obtain the light yield in PA for the 1~MeV source peak.

\begin{figure}[htbp]
  \centering
  \includegraphics[width=.75\textwidth]{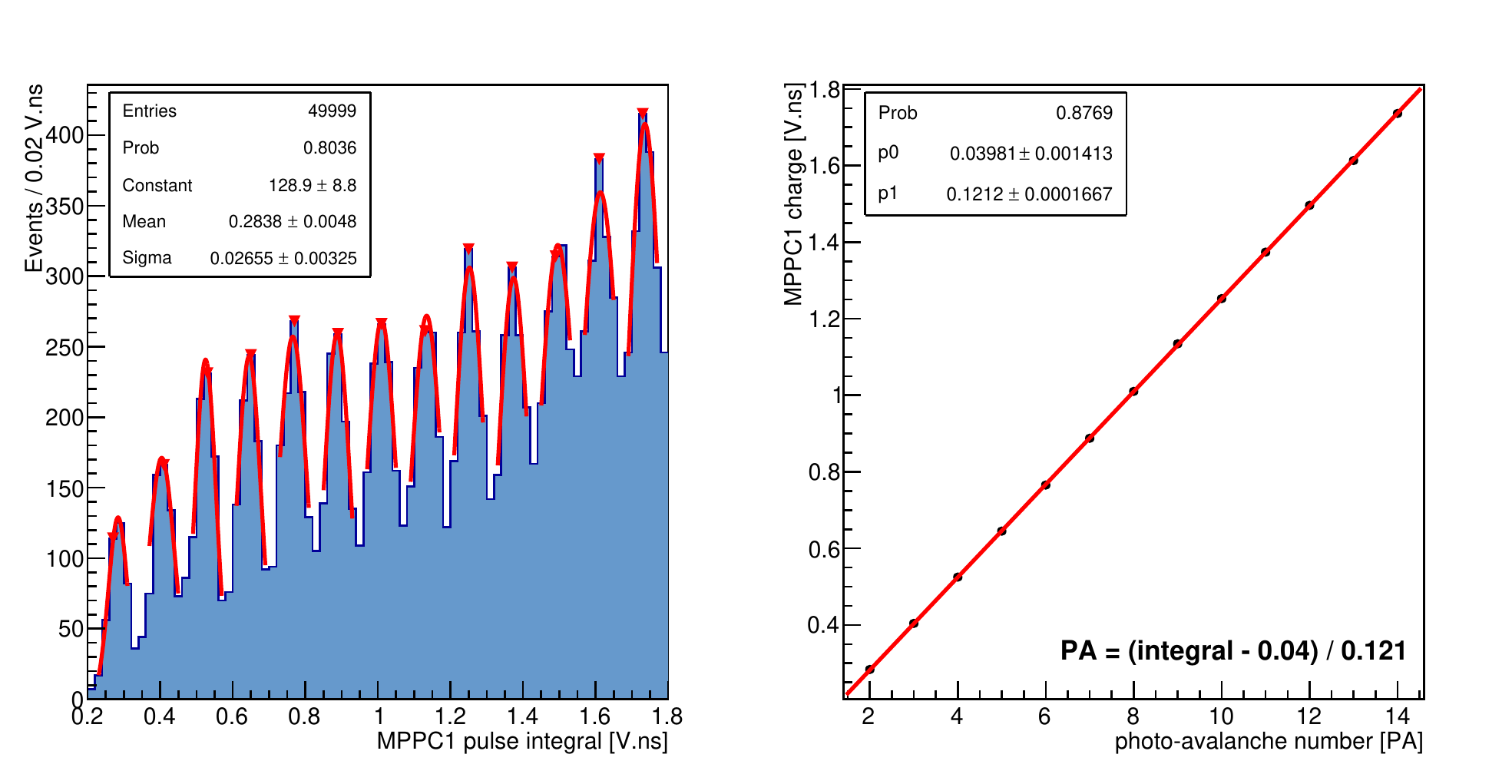}\\
  \includegraphics[width=.75\textwidth]{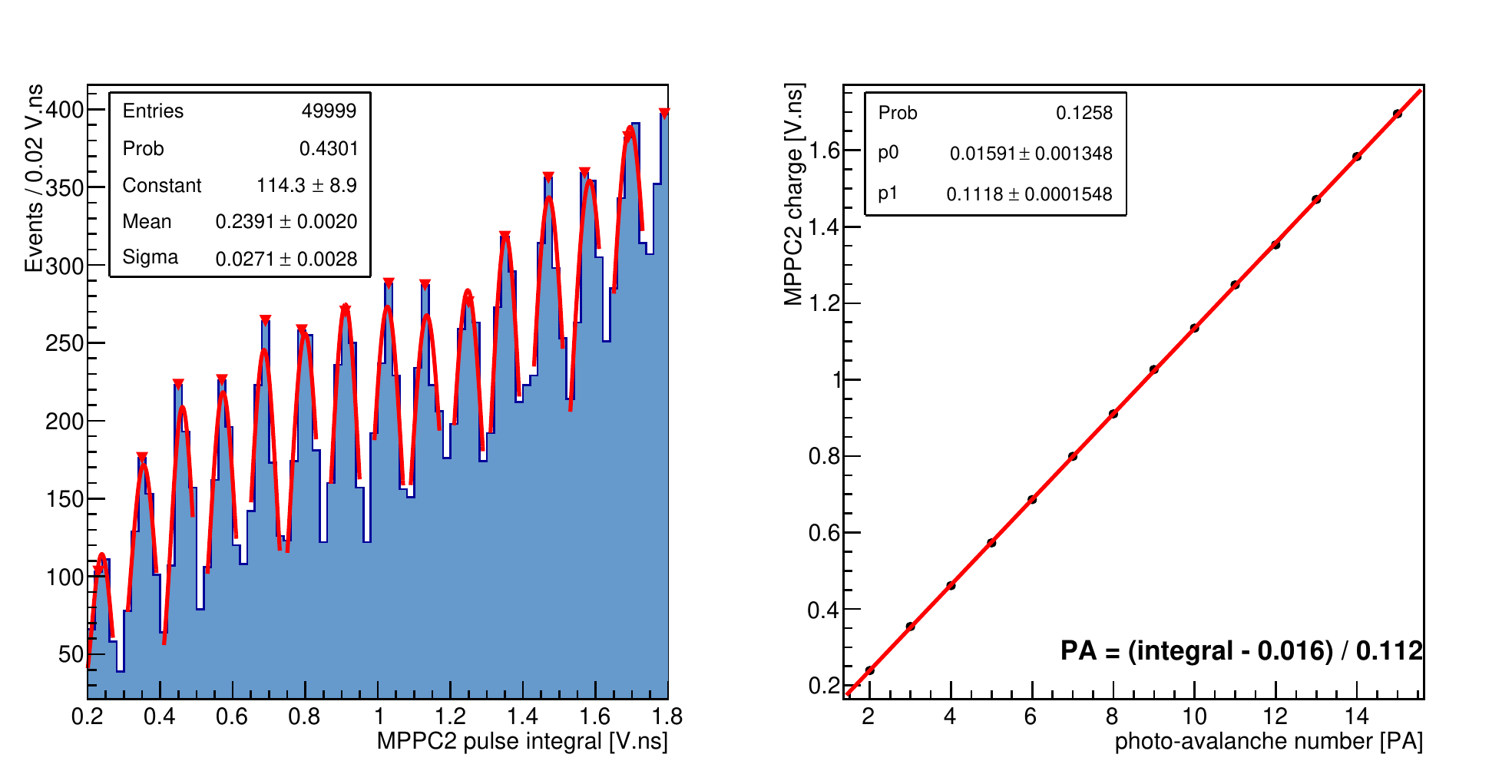}
  \caption{\label{fig:mppc-calib} The left panels show the low energy part of the pedestal subtracted integral spectra with the individual PA fits. The right panels show the linear calibration fit between the integral and the number of~PAs.}
\end{figure}

The first step consists of calibrating the MPPC integral response to a number of PA. To achieve this, the low energy part of the integral spectrum is considered after pedestal subtraction. Using ROOT \cite{root} about 10 individual PA peaks are identified and fitted with a Gaussian function (figure~\ref{fig:mppc-calib} left). Each integral peak corresponds to a number of PA and the relation between the integral and the number of PA is fitted by linear function (figure~\ref{fig:mppc-calib} right). This provides the conversion between the integral of the MPPC pulses and the number of PAs.

\begin{figure}[htbp]
  \centering
  \includegraphics[width=.99\textwidth]{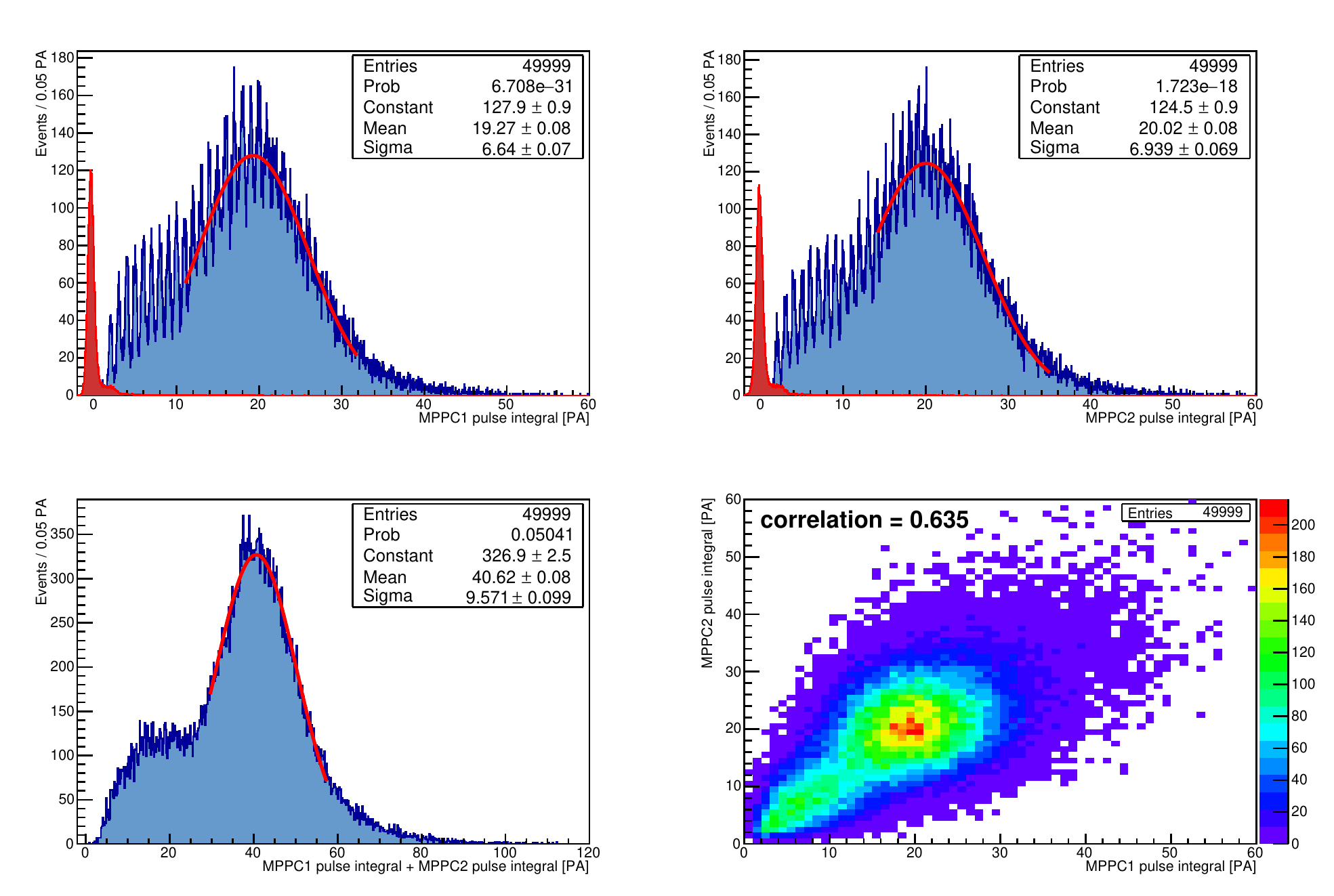}
  \caption{\label{fig:mppc-charges} The top panel shows the calibrated PA spectra for the two MPPCs with the Gaussian fit. The red distribution is the calibrated pedestal spectrum. The bottom left represents the sum of the two MPPC integrals and right shows the correlation between the integrals of the two MPPCs.}
\end{figure}

Figure~\ref{fig:mppc-charges} shows the calibrated integral spectra expressed in PAs of the two MPPCs, their sum and the correlation between the signals. The integral spectra for the individual MPPCs give a similar peak position (here 19.1 and 20.0 PAs) and the linear correlation is over 60~\%. The summed integral spectrum is used to give the final result of the measurement with the 1~MeV peak fitted by a Gaussian function: $N_{PA}$~=~40.5 PA in this example. Subtracting the 17.7~\% cross-talk at 1.5 V over-voltage, it results in a light yield of $N_{PA}$~=~33.3 PA. Given the expected peak eneergy of $\sim$910~keV (section~\ref{subsec:simulation}) it finally corresponds to a light-yield of 36.6 PA/MeV. The stochastic term of the detector energy resolution could then be estimated by $1/\sqrt{N_{PA}}$, which corresponds to 16.5~\% at 1~MeV ($\sqrt{0.91}$ energy losses correction to the resolution) for this cube with only 1 fibre and a double MPPC readout.

This example illustrates the procedure to get the light yield for a given configuration.

\subsection{Measurement uncertainties}
\label{subsec:uncertainties}

The statistical uncertainty of the measurements is negligible since between 30000 and 50000 events are acquired. Indeed the fit of the Gaussian peak is returning a statistical uncertainty of around 0.2~\% on the mean of the function (figure~\ref{fig:mppc-charges}). Therefore the statistical uncertainties in the following will not be mentioned for each measurement.

\begin{figure}[htbp]
  \centering
  \includegraphics[width=.6\textwidth]{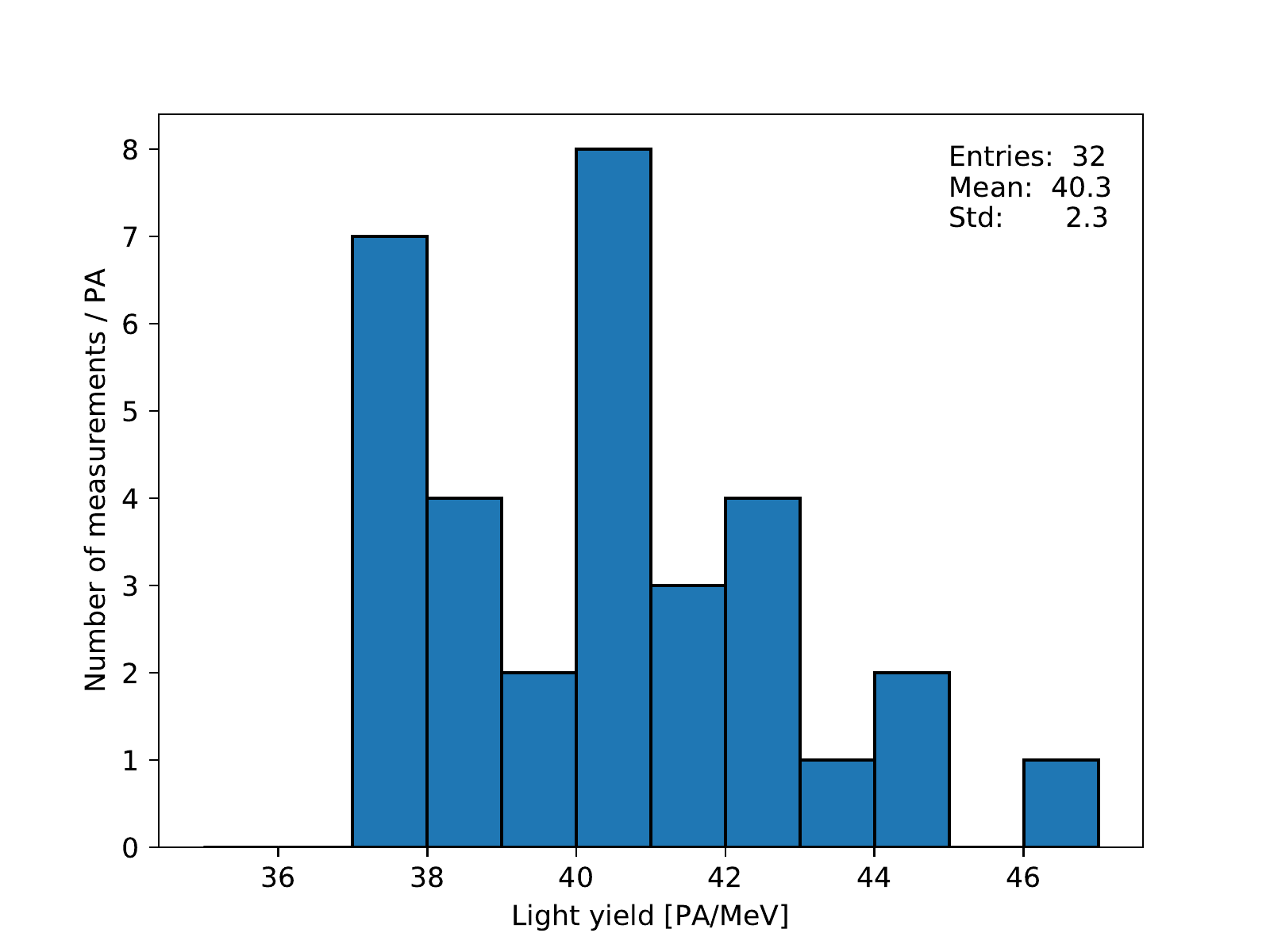}
  \caption{\label{fig:systematics} Light yield measurements performed at different moments for the same reference setup as specified in the text. The standard deviation of this distribution is taken as a measure of the systematic uncertainty.}
\end{figure}

The systematic uncertainty is related to the setting up of a measurement, we have identified several sources of systematic uncertainties that could come from the handling or the positioning of the scintillator cubes and fibres, temperature variations and voltage setting variations. Some of these issues have been addressed separately and will be presented in the next sections. In addition, most of the time this setup is used for comparison between different configurations in order to minimize the systematic uncertainties. This also means that for each test a reference measurement is performed and is in most cases the same (same cube, wrapping and fibre). These measurements have been made over several weeks by different operators at different temperature and voltage settings. Comparing these results leads to an estimation of the total systematic uncertainty. Figure~\ref{fig:systematics} shows the relative variation in PA for 32 of these reference measurements. On average 40.3 PA for the 1~MeV peak with a standard deviation of 2.3 PA is observed, which corresponds to about 5~\%. This value of 5~\% is considered as the systematic uncertainty of the light yield measurements.

\section{Scintillator light collection studies}
\label{sec:cube-studies}

This section presents the studies of the light collection for a single SoLid cube. The influence of the scintillator material, the cube wrapping, the optical fibres and the \up{6}LiF:ZnS(Ag) screen on the light collection is studied.

\subsection{Plastic scintillator production and cleaning}
\label{subsec:material}

For the SoLid Phase~1 detector, the scintillator cube machining has been improved to obtain a better cube surface quality. Polishing the 12800 cubes needed for the experiment would improve the light yield further, but this was not cost effective. Therefore we focussed on optimizing surface quality after machining. In order to estimate the quality of the machining we measured with a roughness meter the surface roughness average (R\down{a}). For SM1 cubes it was around 0.45~\si{\micro\meter} compared to 0.04~\si{\micro\meter} for the new cubes. This increased the light yield by 10~\%.

In order to prevent the scintillator from heating, a lubricant is used during the machining. This leaves a grease film on the surface of all the cubes. As a reference measurement, the light yield of a cube was measured directly after machining, hence before cleaning. This gave a light yield of 35 PA/MeV. The cubes were then cleaned by hand in a soap solution at room temperature, rinsed with demineralised water and left to dry in the air or with tissues. Two other cleaning methods were tested: cold ultra-sonic bath and the same cleaning method as before but using a nylon brush to better clean the grooves. The cold ultrasonic bath was not efficient, increasing the light yield by only 5~\%. All other cleaning methods were equivalent increasing the light yield by 25~\% as long as enough soap was used and the cubes were well rinsed.

\subsection{Cube wrapping material}
\label{subsec:wrapping}

Teflon (or PTFE) is known to be one of the best reflective materials for scintillation light. A SoLid cube was wrapped with 0.2 mm thick Teflon tape (80~\si{\gram\per\square\meter}) and tested. The result of the measurement with Teflon leads to the best light yield measured in this configuration, giving 44 PA/MeV. However, the wrapping of cubes with Teflon tape, leaving a hole for the fibre and avoiding extra layers for electron energy loss is time consuming and error prone. Since the SoLid Phase~1 contains 12800 cubes, Teflon tape was excluded for practical reasons. Nevertheless, this test provides a good reference to select appropriate wrapping material.

\begin{figure}[htbp]
  \centering
  \includegraphics[width=.45\textwidth]{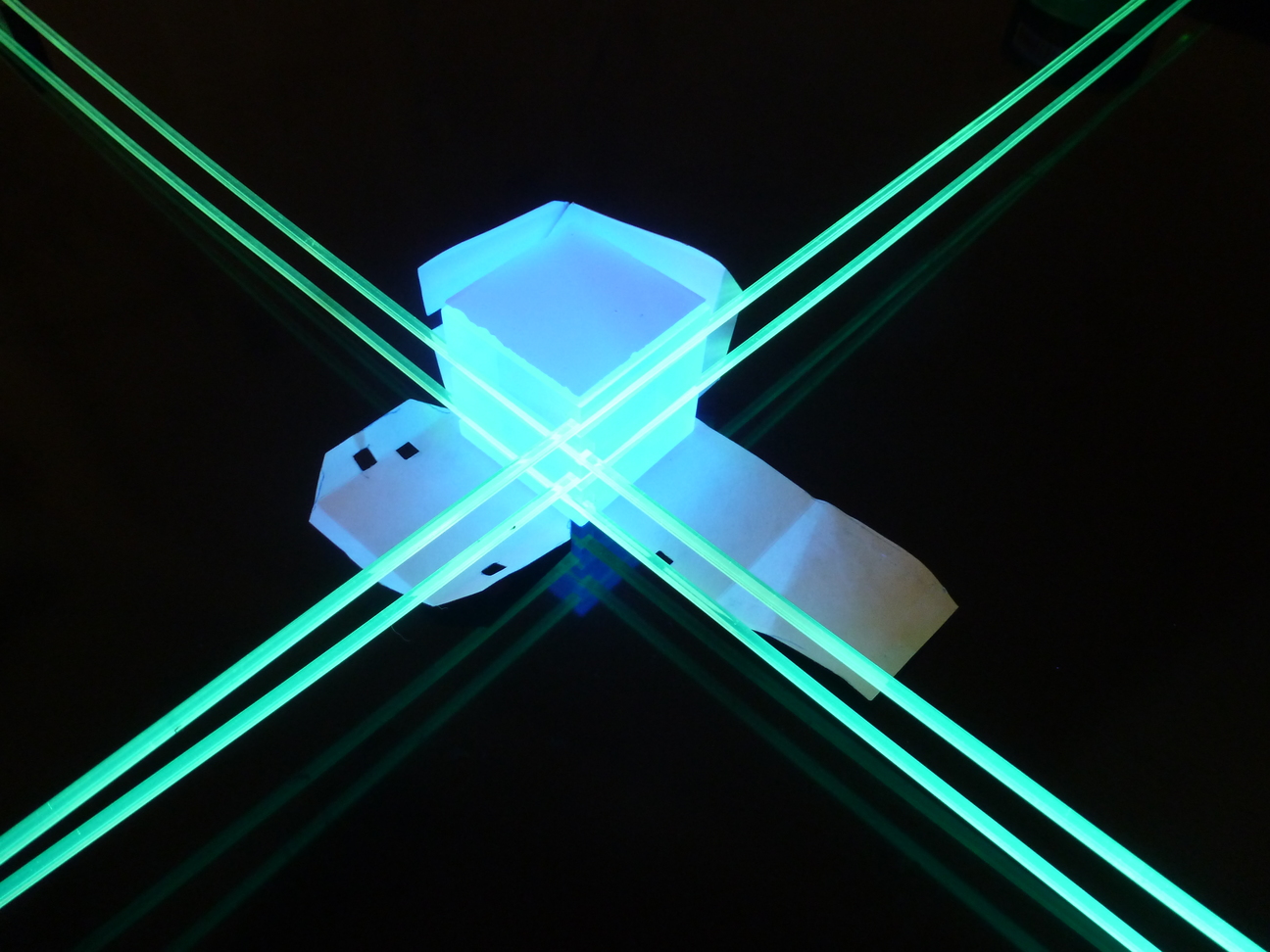}
  \caption{\label{fig:cube-uv-picture} Picture of a SoLid cube with its Tyvek wrapping opened, a \up{6}LiF:ZnS(Ag) sheet on top of the cube and four optical fibres. The cube is illuminated with UV light to highlight the detector components. On this prototype cube, the fibres were going along two faces of the cube instead of four faces in the final design.}
\end{figure}

Tyvek is another very good candidate for reflecting scintillation light. It is also much more convenient to use as wrapping for the cubes since it is possible to cut and pre-fold a pattern using press techniques. This is shown in figure~\ref{fig:cube-uv-picture} where the Tyvek wrapping is unfolded around a cube. This material was already used for the SM1 detector but, as discussed in section~\ref{subsec:simulation}, the Tyvek used at that time was not the thickest possible. Indeed for cubes assembled in the detector plane, the surrounding Tyvek layers from other cubes contributed to an increase of the light yield compared to a single cube. To quantify this effect, up to four layers of Tyvek wrapping have been added successively around a PVT cube. The second layer improved the light yield by about 20~\%, the third one gave an extra 10~\% with respect to two layers while the fourth one had no additional effect. For the construction of the SoLid Phase~1 detector it is not convenient to use several layers of wrapping around each cube so we have selected the thickest Tyvek from DuPont\up{\texttrademark} (1082D as presented in section~\ref{subsec:simulation}). A light yield of 36.7 PA/MeV was measured for this Tyvek compared to 33.6 PA/MeV for the Tyvek used in the SM1 detector. This is an improvement of 10~\% for a single cube. Although this is a 15~\% lower light yield than Teflon, it was the best material found taking into account construction constraints.

\subsection{Optical fibres}
\label{subsec:optical-fibres}

\begin{figure}[htbp]
  \centering
  \includegraphics[width=.55\textwidth]{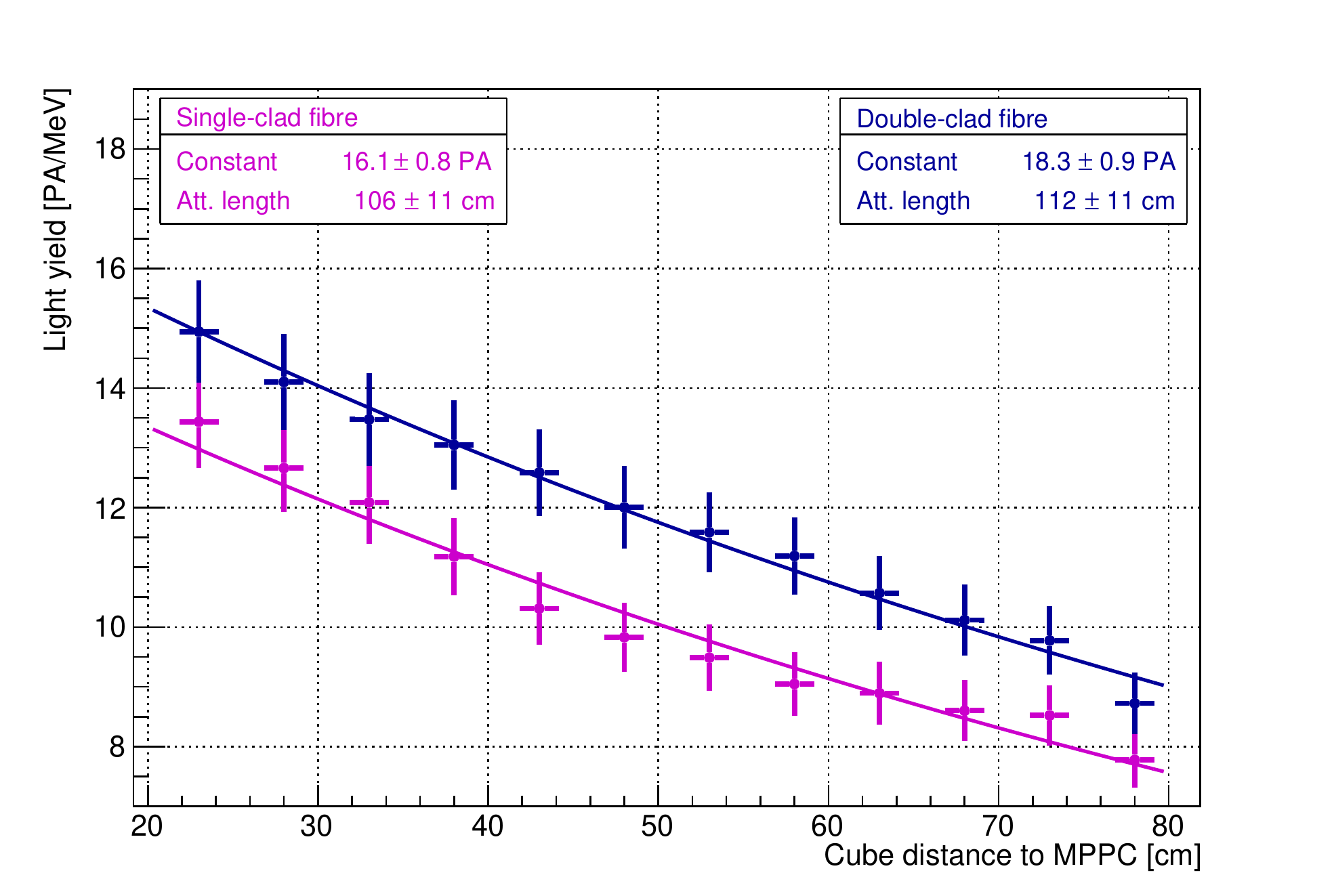}
  \caption{\label{fig:attenuation-single-double} Comparison of single-clad and double-clad BCF-91A optical fibre from Saint-Gobain used in the SM1 prototype and SoLid Phase~1 detector. The measurements were performed at the same time with a single cube and a single MPPC readout per fibre. The uncertainties corresponds to a 5~\% systematic uncertainty on the Y-axis and a 1~cm positioning precision on the X-axis.}
\end{figure}

When the SM1 detector was constructed only single-clad squared fibres were available. However for the Phase~1 detector, Saint-Gobain was able to produce double-clad squared fibres. The test bench has been prepared with one single-clad fibre used in the SM1 detector and one double-clad fibre used for the Phase~1 detector going through the same cube, at the same time, to be able to compare both. The two MPPCs are each connected to one of the fibres and the other extremity is left free. The Fresnel reflection at the fibre-air interface of about 5~\% was ignored in the further analysis. The assembly is mounted on the rail to allow the cube translation along the fibres. The result of 12 measurements along the fibres at different cube positions is presented in figure~\ref{fig:attenuation-single-double}.

The exponential decay fit of the light yield as a function of the distance shows that about 15~\% more light is trapped by the double-clad fibre (`Constant' parameter of the fit). This is less than expected since up to 60~\% higher trapping efficiency could be theoretically obtained with double-clad fibres compared to single-clad fibres. Similar small beneficial effect has been observed when comparing squared fibers in other measurements (see \cite{fibers-alfa} for example).

The attenuation length for single and double-clad fibres are respectively measured to 106$\pm$11 and 112$\pm$11~cm. Thus we don't observe difference in the attenuation lengths for both type of fibres. These values are quite short for attenuation lengths. Light transport in optical fibres is based on different modes, depending on the distance of the excitation point from the photodetector. Helical and cladding modes contribute to a fast dropping attenuation length on the first tens of cm, while the core (meridional) modes produce the long component. In the SoLid detector configuration we are then sensitive to the fast dropping component.

Varying the fit range on the data shown in figure~\ref{fig:attenuation-single-double} gives a systematic uncertainty for these measurements, resulting in a change in light yield by 10~\% and the attenuation length by 20~\%. These results show that double-clad fibres give an improvement in term of light yield compared to single-clad fibres. Therefore, the double-clad fibres are used for the SoLid Phase~1 experiment.

\subsection{\up{6}LiF:ZnS(Ag) neutron screens}
\label{subsec:zns}

Three types of NS produced at different times are used for the construction of the SoLid Phase~1 detector. The first two generations were fragile so a third generation was produced with a less fragile substrate as backing.

PVT scintillator light yield measurements with the different types of NS have been performed. Two NS per PVT cube were used, similar to the SoLid Phase 1 detector design. A light yield of 27.2~PA/MeV for the oldest generation of NS, 29.9~PA/MeV for the second generation and 29.8~PA/MeV for the NS with a less fragile backing have been measured. The oldest generation results in about 10~\% lower light yield. The three types of screens had to be used for the construction of the SoLid Phase~1 detector. Because of this lower light yield and greater fragility, the oldest \up{6}LiF:ZnS(Ag) scintillators are used in the external layer of the detector planes where neutron detection efficiency is lower because of edge effects.

When neutrons interact in the NS the emitted light will have to go through the plastic scintillator before being trapped in the fibre. Since the sensitivity of the light yield to the wrapping material is large, an important impact of having a NS between the cube and its wrapping is expected. For the SM1 prototype only one NS per cube was used. For the SoLid Phase~1 detector, two NS will be used since simulation studies have shown that neutron detection efficiency could significantly increase, reducing at the same time the neutron capture time. One of the screens will be oriented perpendicular to the antineutrino direction to increase efficiency. The second NS will pass along a fibre between the PVT scintillator and the Tyvek (section~\ref{subsec:cube-design}).

To check this hypothesis, the light yield measurements were performed for a cube wrapped with SoLid Phase~1 Tyvek and either one fibre without NS, or with one NS sheet on a face without fibre, or with the same NS on a face where the fibre is going through the cube. For these three configurations respectively 33.6, 30.6 and 29.7 PA/MeV were measured. The first drop of about 9~\% confirms that adding \up{6}LiF ZnS:Ag decreases the PVT light yield. The loss is then only $\sim$3~\% when the surface of one NS is parallel to the fibre. This effect is close to our systematic error but is significant.

In conclusion for the SoLid Phase~1 detector, the plastic scintillator light loss due to the NS will be limited to about 12~\% thanks to the fact that one of the two NS sheets will be placed along an optical fibre between the PVT scintillator and the Tyvek instead of covering a face of the cube where no fibre is going through.


\section{Detector configuration studies}
\label{sec:detector-studies}

In this section the detector design and configuration will be studied for what concerns the light yield of individual scintillator cubes.

\subsection{Position of the fibres in the scintillator cube}
\label{subsec:cube-design}

For the SM1 cubes, squared grooves at the surface of the cube were holding the optical fibres (figure~\ref{fig:cube-designs} left). This design was relatively easy to machine and allowed for easy detector assembly. For the SoLid Phase~1 cubes a design was considered with the fibre going through the core of the cube to have more scintillating material surrounding the fibre. Cubes with circular holes drilled through the scintillator were tested and resulted in a 10~\% increase in light yield. However, when considering the machining time, the cost and a possible heating damage to the scintillator during drilling, this design solution was not selected. Several positions for the surface grooves where then considered, but the actual position of the grooves turned out not to be important for the light yield. Hence the position of the grooves was driven by the detector mechanical design. The scintillator cube design has been optimized with four grooves on four faces with 2.5 mm spacing as shown in figure~\ref{fig:cube-designs} right. The four fibres remain in the 16$\times$16 cubes plane to allow the stacking of the detector planes along the neutrino direction.

\subsection{Number of fibres per scintillator cube and readout scheme}
\label{subsec:fibres-per-cube}

The SM1 detector was limited to two fibres per cube with a single readout. One potential optimization would be to have a double readout per fibre. Another option would be to have 4 fibres with a single readout. Both options result in a higher light yield. To decide which option is best a comparison was performed between single and double readout of a fibre. The test bench does not allow for reading out four fibres. Therefore, the measurement consisted of measuring the light yield of one fibre read out by one MPPC on one end and with or without a mirror at the other end of the fibre. The result is then extrapolated to the detector configuration. Because of the reflection at the other extremity of the fibre on air \footnote{The free extremity of the fibre in air produces a reflection of light of about 5~\%.} or on the mirror, the result depends on the position of the cube along the fibre (section~\ref{subsec:fibres-reflector}). Since this measurement could not be done for all the positions along the fibre, the cube was placed in the middle of the fibre ($\sim$45 cm from MPPC and mirror or free extremity). This position should correspond to an average response of the detector. The materials used were selected based on the studies in section~\ref{sec:detector-studies}. With mirror a light-yield of 25.3~PA/MeV has been measured and 15.9~PA/MeV without mirror. The addition of the mirror represents then an improvement of 60~\%. For a double readout per fibre, the light yield of a fibre without mirror would be doubled.

For two fibres with double readout, we would therefore obtain 63.6~PA/MeV. For four fibres with single readout and a mirror, the light yield would be 101.2~PA/MeV. Based on this estimation, the latter configuration would be preferred. However, putting more fibres in the cube will reduce the amount of light collected per fibre. To quantify the reduction of the light yield due to the presence of other fibres an additional measurement is performed. A reference fibre is inserted in the cube with a double readout. Additional fibres are then inserted one by one into the cube and the light yield for the first fibre is measured. Since the cube is already machined with four grooves this measurement cannot take into account a possible light reduction produced by the grooves themselves. The result is presented in table~\ref{tab:number-of-fibres}. Each new fibre that is introduced takes on average $\sim$15~\% of the light from the first one. The third row in the table~\ref{tab:number-of-fibres} shows that $\sim$16~\% less light is collected per fibre with the two fibres design and 40~\% less light per fibre in the four fibres design. With this reduction, the estimated light yield for the two fibres with double readout is 53.4~PA/MeV compared to 60.7~PA/MeV for the four fibres with single readout and a mirror. Hence, the configuration with four fibres with single readout performs 15~\% better in terms of light yield. Based on these studies, the four fibres configuration with single readout was adopted for the SoLid Phase~1 detector.

\begin{table}[htbp]
\centering
\caption{\label{tab:number-of-fibres} Impact of the number of double-clad optical fibres inserted in the plastic scintillator grooves on the light yield of the first fibre with double readout. Adding other fibres decrease the light-yield per fibre but increases the total light-yield.}
\smallskip
\begin{tabular}{|l|c|c|c|c|}
  \hline
  \textbf{Number of fibres} & \textbf{1} & \textbf{2} & \textbf{3} & \textbf{4} \\
  \hline
  light yield for the first fibre [PA/MeV] & 33.1 & 27.6 & 24.1 & 19.8 \\
  \hline
  Variation to previous [\%] & - & -16.4 & -12.9 & -17.6\\
  \hline
  Variation to 1 fibre [\%] & - & -16.4 & -27.2 & -40.0\\
  \hline
  Total light yield [PA/MeV] & 33.1 & 55.2 & 72.3 & 79.2\\
  \hline
\end{tabular}
\end{table}

As the last line of table~\ref{tab:number-of-fibres} is showing, even if the light-yield per fibre is decreasing, the total light-yield per cube is still increasing. These measurements give a comparison of the total light expected in the two and four fibres per cube configurations and we can expect an increase of 43~\% in the light collection compared to the SM1 like configuration with 2 fibres. This is another argument to prefer the four fibres per cube with single readout design.

Finally, to verify whether the four fibres are collecting the same amount of light, we then have measured the light yield four times, moving the same fibre each time in a different groove. These four measurements give a light yield that is consistent within 4~\%, which is smaller than the systematic uncertainty. Hence the location of the fibre does not matter in terms of light yield. We also rotated the cube along the fibre direction to check different faces of the scintillator cube. We do not observe differences in all these measurements either. These tens of measurements indicate that the scintillation light is uniformly distributed in the scintillator volume, confirming the results in section~\ref{subsec:trigger} where the response for localized electron interactions and gamma interactions in the whole scintillator volume are compared.

\subsection{Spatial freedom of the optical fibre}
\label{subsec:optical-fibres-configuration}

As already mentioned, the size of the grooves in the SoLid scintillator cubes is 5$\times$5~\si{\square\mm} to hold the 3$\times$3~\si{\square\mm} squared fibres. The relatively large grooves facilitate the insertion of the fibres once a detection plane is assembled. As consequence the fibre has the possibility to move in the grooves. We have measured the effect of the positions of the fibre with respect to the cube to quantify the reproducibility of the results. The maximal observed effect was the rotation of the fibre. The four measurements at different rotation angles vary less than 4~\% as can be seen in table~\ref{tab:fibre-rotation}. This effect is within the systematics uncertainty, which implies that the position and orientation of the fibre in the groove have no effect on the light collected by the fibre. It is worth mentioning that these large rotation angles are not possible in the SoLid detector design since the fibres are held by 3D printed connectors at the two extremities to hold them in position in the planes.

\begin{table}[htbp]
\centering
\caption{\label{tab:fibre-rotation} Effect rotating the fibre in the groove. Angle~\si{\ang{0}} means that three faces of the fibre are parallel to the three faces of the cube groove.}
\smallskip
\begin{tabular}{|l|c|c|c|c|}
  \hline
  Fibre angle &~\si{\ang{0}} &~\si{\ang{45}} &~\si{\ang{90}} &~\si{\ang{-45}} \\
  \hline
  light yield [PA/MeV] & 38.6 & 40.0 & 39.8 & 39.5 \\
  \hline
  Variation to~\si{\ang{0}} [\%] & - & 3.6 & 3.1 & 2.3 \\
  \hline
\end{tabular}
\end{table}

\subsection{Reflector at the end of the optical fibre}
\label{subsec:fibres-reflector}

\begin{figure}[htbp]
  \centering
  \includegraphics[width=.55\textwidth]{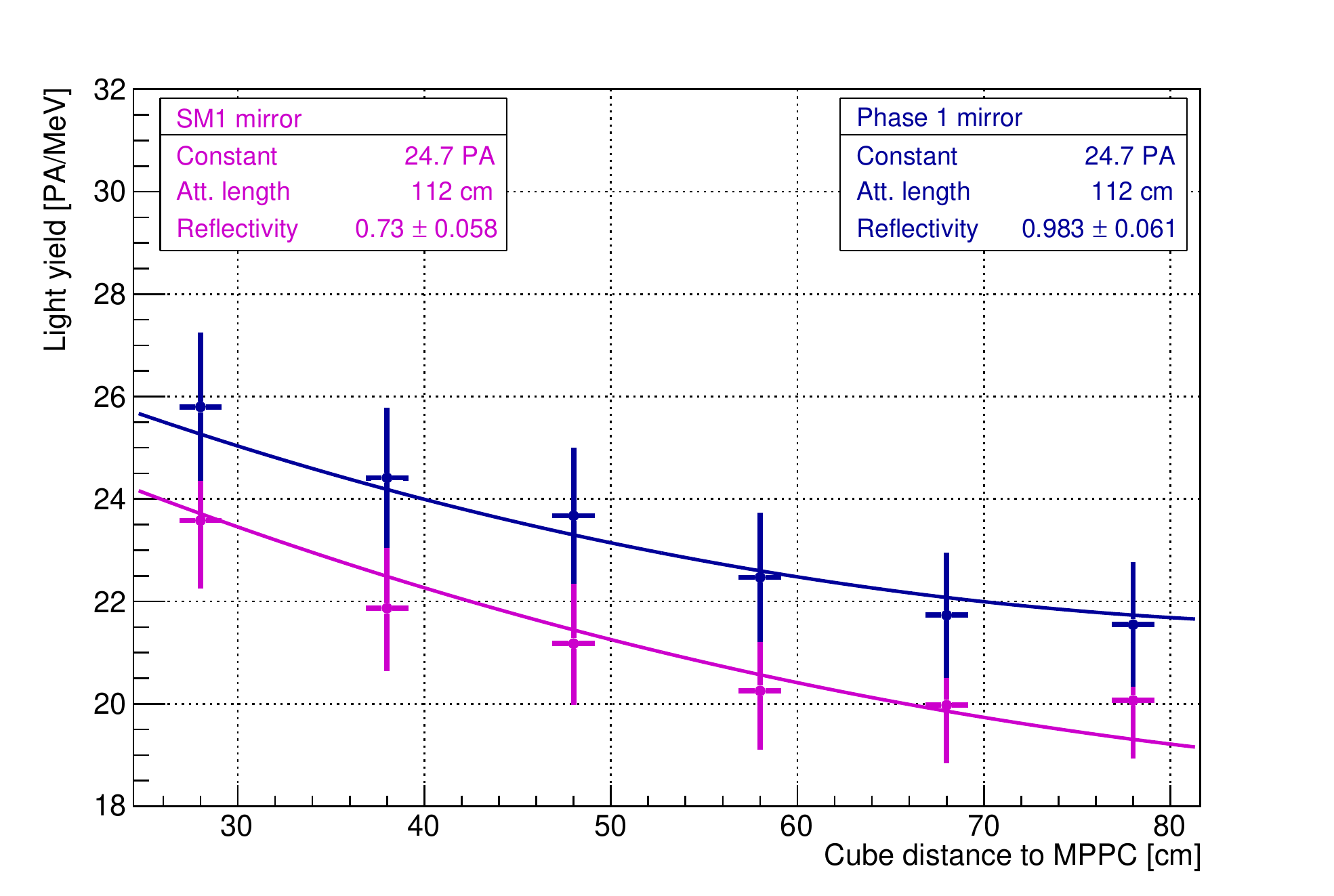}
  \caption{\label{fig:attenuation-mirror} Comparison of the aluminium sticker mirror used for the SM1 fibres and the 70~\si{\micro\meter} aluminised mylar film mirrors used for the fibres in the SoLid Phase~1 detector. The cube and fibre used for the measurement is the same in both cases.}
\end{figure}

In section~\ref{subsec:cube-design} the impact of using a mirror at one end of the fibres has been shown. Therefore we investigated the impact of the type of mirror. For the SM1 fibres, an aluminium sticker mirror was used. We have explored different other options and tested aluminised mylar film. The aluminium has a standard thickness of $\sim$200 nm. Several thicknesses for the mylar were possible, but showed no differences in light yield. A mylar thickness of 70~\si{\micro\meter} was selected for its rigidity, which is more convenient when inserting the end of the fibre in the 3D printed connectors. We compared the mirror used in SM1 and the aluminised mylar mirror using the same cube and the same fibre with a single MPPC readout. We measured the light yield for six distances along the fibre in both cases. The result is presented in figure~\ref{fig:attenuation-mirror}. The function used for the fit is given by equation~\ref{eqn:fit-mirror}, which is taking into account the reflection at the end of the fibre with the mirror.
\begin{equation}
  f(x) = C~(e^{-x/L_{att}} + R~e^{-(2\times L_{fibre}~-~x)/L_{att}})
  \label{eqn:fit-mirror}
\end{equation}
where $C$ is a normalisation coefficient, $L_{att}$ is the attenuation length in cm, $R$ the light reflection coefficient of the mirror and $L_{fibre}$ is the total length of the fibre, which is 92.2~cm.

In order to compare only the reflection coefficient, the normalisation coefficient and the attenuation length are fixed to 24.7~PA/MeV and 112~cm, respectively, as determined from previous measurements. We find that the SM1 mirror has a reflection coefficient of 73 $\pm$ 6~\% while it is 98 $\pm$ 6~\% for the other mirror. These fitted values are certainly optimistic for mirrors but we are only concerned about relative gain between both resulting light-yields. There is also correlation between attenuation length and reflectivity measured this way. Consequently the aluminised mylar mirrors have been selected for the SoLid Phase~1 detector. The effect on the total light yield per cube depends on its position along the fibre because of attenuation. For example this mirror would produce an increase of light yield per fibre of 5~\% for the cube farthest to the mirror, 7~\% for a cube at the centre and 11~\% for the cube closest to the mirror.

\subsection{Impact of neighbouring cubes}
\label{subsec:neighbouring-cubes}

\begin{figure}[htbp]
  \centering
  \includegraphics[width=.55\textwidth]{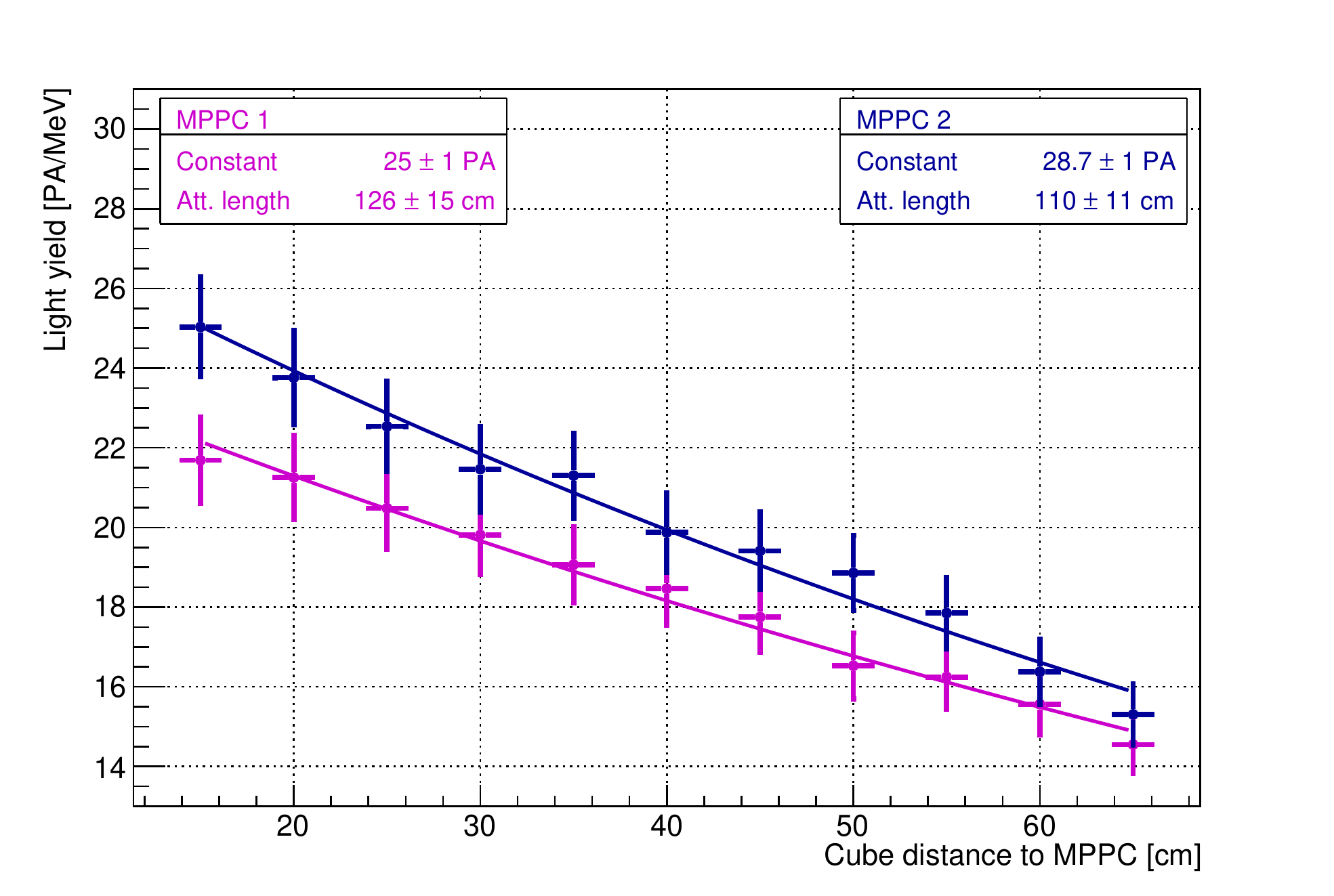}
  \caption{\label{fig:attenuation-16cubes} Attenuation along a double-clad fibre going through 16 cubes with double readout. The MPPC light yield is corrected for the individual cube light yield that is different for each cube along the fibre.}
\end{figure}

In the SoLid Phase~1 detector the cube and fibre environment is different than that in the test bench. Indeed the fibres will be surrounded by scintillator cubes along their full length. This could have an impact on the light yield for a single cube or on the attenuation length. We performed a test with 16 cubes positioned along one double-clad fibre, which is read out by two MPPCs. Considering a cube close to the center of the fibre, we observe an increase of the light yield of 12~\% compared to the same measurement where only one cube was positioned along the fibre (figure~\ref{fig:attenuation-single-double}). Since the \up{207}Bi source and the trigger system are free to move along the fibre the light yield of each of the 16 cubes was measured. The measurements are normalised to the sum of the two MPPC signals for each cube to cancel the potential effect of a different response for the cubes. The result of the attenuation measurement for the individual MPPC signals after correction is presented in figure~\ref{fig:attenuation-16cubes}. The attenuation length seems to increase a bit although the uncertainty is quite large. The difference between the two MPPCs is partially due to the difference in breakdown voltages. This measurement implies that the light yield will be better in the real detector where 16$\times$16 cubes are assembled in planes compared to our test bench studies.

A second effect we measure in this test, is the light escaping to neighbouring cubes. This optical cross-talk could for instance come from light going through the Tyvek, but is more likely to come from leaks through the holes in the Tyvek where the fibres pass. A second cube is placed next to the one interfaced with the calibration source. Two fibres were put through these two cubes perpendicularly to the fibre going through the 16 cubes. Some light has been observed in the neighbouring cube with a peak in the integral spectrum between 1 and 2 PA. After calibrating the light collected by this cube with the \up{207}Bi source, we conclude that in 90~\% of the cases we record less than 10~\% ($<$~100~keV) of the light in the cube next to the source. The correlation with the integral spectrum of the cube with the source is weak ($<$~0.2) but this might be due to the low number of PA measured. Optical cross-talk should not affect the energy reconstruction for the SoLid experiment since it is very low and the four channel readout per cube will allow for distinguishing the different light origins. We have performed the same test with the next to next cube but no light excess was visible.

\section{Summary of the light yield improvements for the SoLid Phase~1 detector}
\label{sec:comparison-sm1-solid}

Table~\ref{tab:light yield-summary} summarizes all the improvements for the light yield of the Phase~1 detector based on the studies presented in this article. The overall light yield improvement is expected to be of about a factor 2.1 compared to SM1 design.

\begin{table}[htbp]
\centering
\caption{\label{tab:light yield-summary} Summary of all the light yield effects for the SoLid Phase~1 detector compared to the SM1 detector.}
\smallskip
\begin{tabular}{|l|c|c|c|}
  \hline
  \textbf{Detector component} & \textbf{SM1} & \textbf{SoLid Phase~1} & \textbf{Relative light yield effect} \\
  \hline
  Cube machining & R\down{a} = 0.45~\si{\micro\meter} & R\down{a} = 0.04~\si{\micro\meter} & + 10~\% \\
  Cube wrapping & 75~\si{\gram\per\square\meter} & 105~\si{\gram\per\square\meter} & + 10~\% \\
  Optical fibre & single-clad & double-clad & + 15~\% \\
  Number of neutron screens & 1 screen & 2 screens & - 3~\% \\
  Number of fibres & 2 per cube & 4 per cube & + 43~\% \\
  Mirror & aluminium & aluminised mylar & + 7~\% \\
  \hline
  \multicolumn{3}{|l|}{Overall expected gain} & $\times$ 2.1\\
  \hline
\end{tabular}
\end{table}

\begin{figure}[htbp]
  \centering
  \includegraphics[width=.7\textwidth]{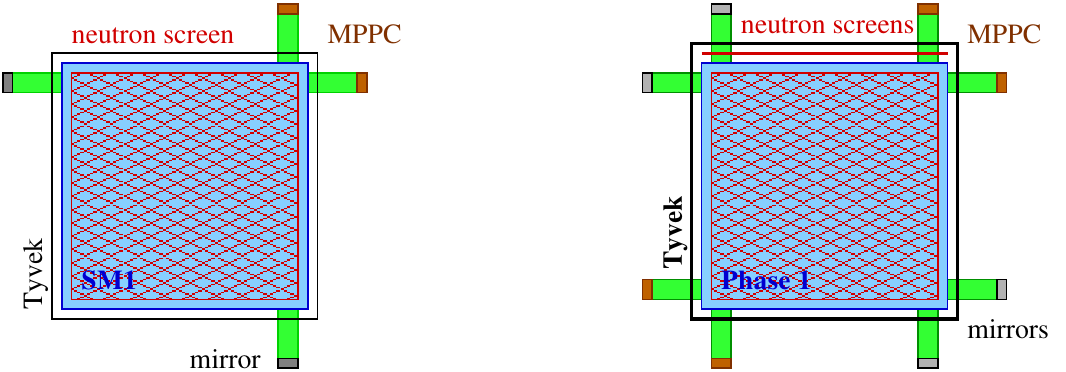}
  \caption{\label{fig:measurement-sm1-phase1} Illustration of the measurements done in the SM1 (left) and Phase 1 (right) SoLid detector configurations to validate the material improvements for the light-yield. They concern the cube machining, Tyvek wrapping, number and position of neutron screens, optical fibres type and numbers and aluminum mirror. The fibre lengths have been reduced for the drawing only.}
\end{figure}

In order to validate all these improvements for the SoLid Phase~1 detector design together, we performed two more measurements in a configuration as close as possible to either the SM1 or Phase~1 design (see figure~\ref{fig:measurement-sm1-phase1}). For the SM1 configuration we have used an SM1 cube with one SM1 neutron screen, SM1 Tyvek, two single-clad fibres with each an MPPC on one end and an SM1 mirror at the other end. For the SoLid Phase~1 configuration, we have used a Phase~1 cube with two Phase~1 NS, Phase~1 Tyvek, four double-clad fibres with each an MPPC on one end and an aluminised mylar mirrors on the other end. Since the prototype amplifier board has only three channels, the measurement was repeated for the four fibres case changing only position of the MPPC for the two measurements and the results were summed to provide the Phase 1 light-yield.

For the SM1 configuration we obtain a total cube light yield of 18.6~PA/MeV and for the SoLid Phase~1 configuration 51.6~PA/MeV. This is an improvement of almost a factor 2.8 in the light yield for one cube of the new detector. This is better than the prediction computed in Table~\ref{tab:light yield-summary} which was a simple cumulation of improvements and was not taking into account all possible effects and the inter-dependence of effects. With this light yield the energy resolution target of $\sigma_E/E = 14$~\% at 1~MeV should be achieved for the SoLid experiment.

The measured light yield for the SM1 configuration is almost 30~\% lower than the observed value for the real SM1 detector, which was 24~PA/MeV \cite{sm1}. This difference is certainly dominated by the impact of neighbouring cubes in the real SM1 detector (sections~\ref{subsec:wrapping} and \ref{subsec:neighbouring-cubes}). To lesser extent, it could be due to the different set-up or electronics. The result for the Phase~1 cube is in agreement with the calculation presented in section~\ref{subsec:fibres-per-cube} where 60.7~PA/MeV was expected for four fibres in the same cube but without the two neutron screens. Using the results presented in section~\ref{subsec:zns} adding two NS would reduce the light yield to 53.6~PA/MeV, which is in very good agreement with 51.6~PA/MeV measured in this last test given the systematic uncertainty of 5~\%.

\begin{figure}[htbp]
  \centering
  \includegraphics[width=.49\textwidth]{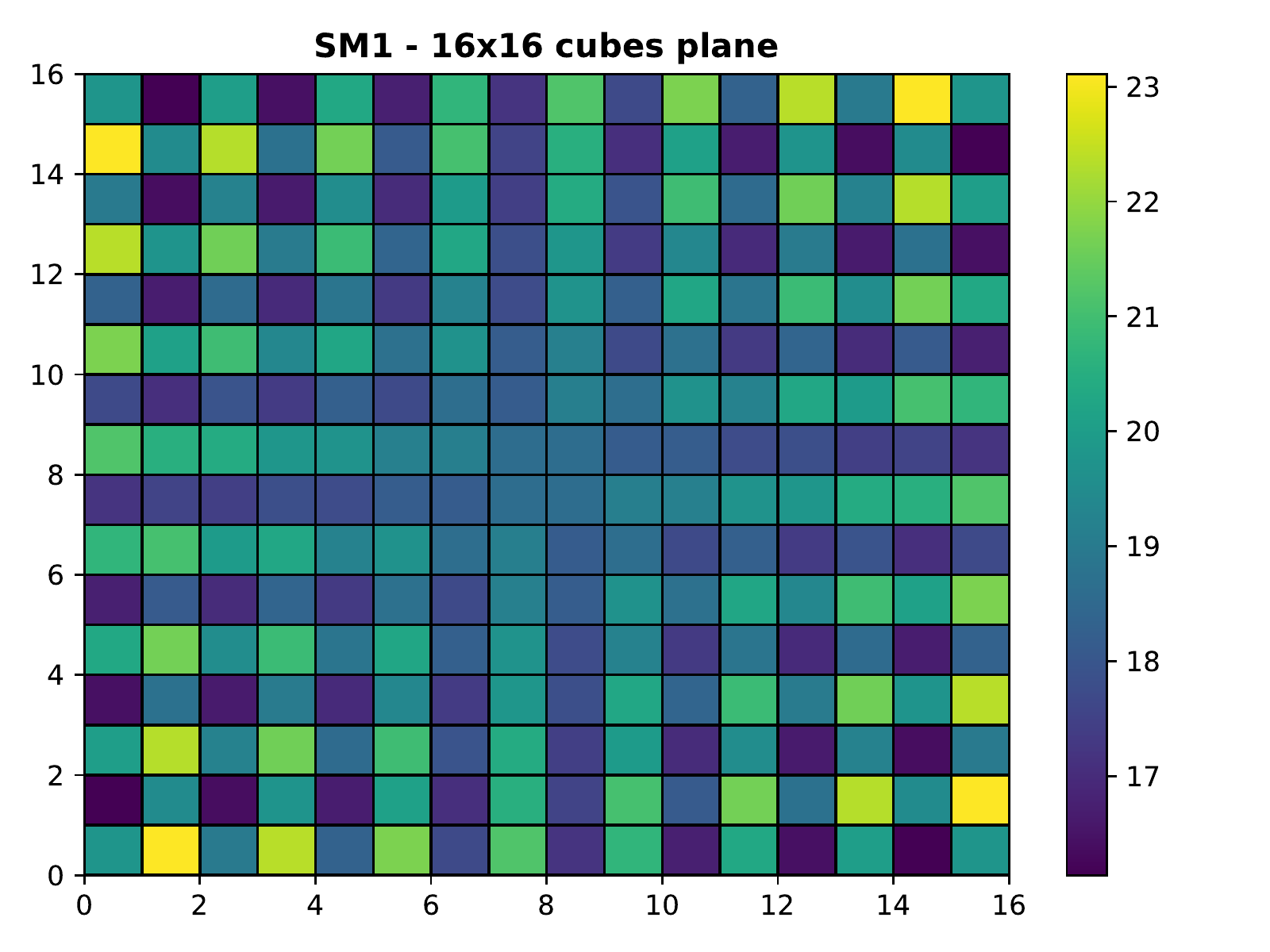}~~
  \includegraphics[width=.49\textwidth]{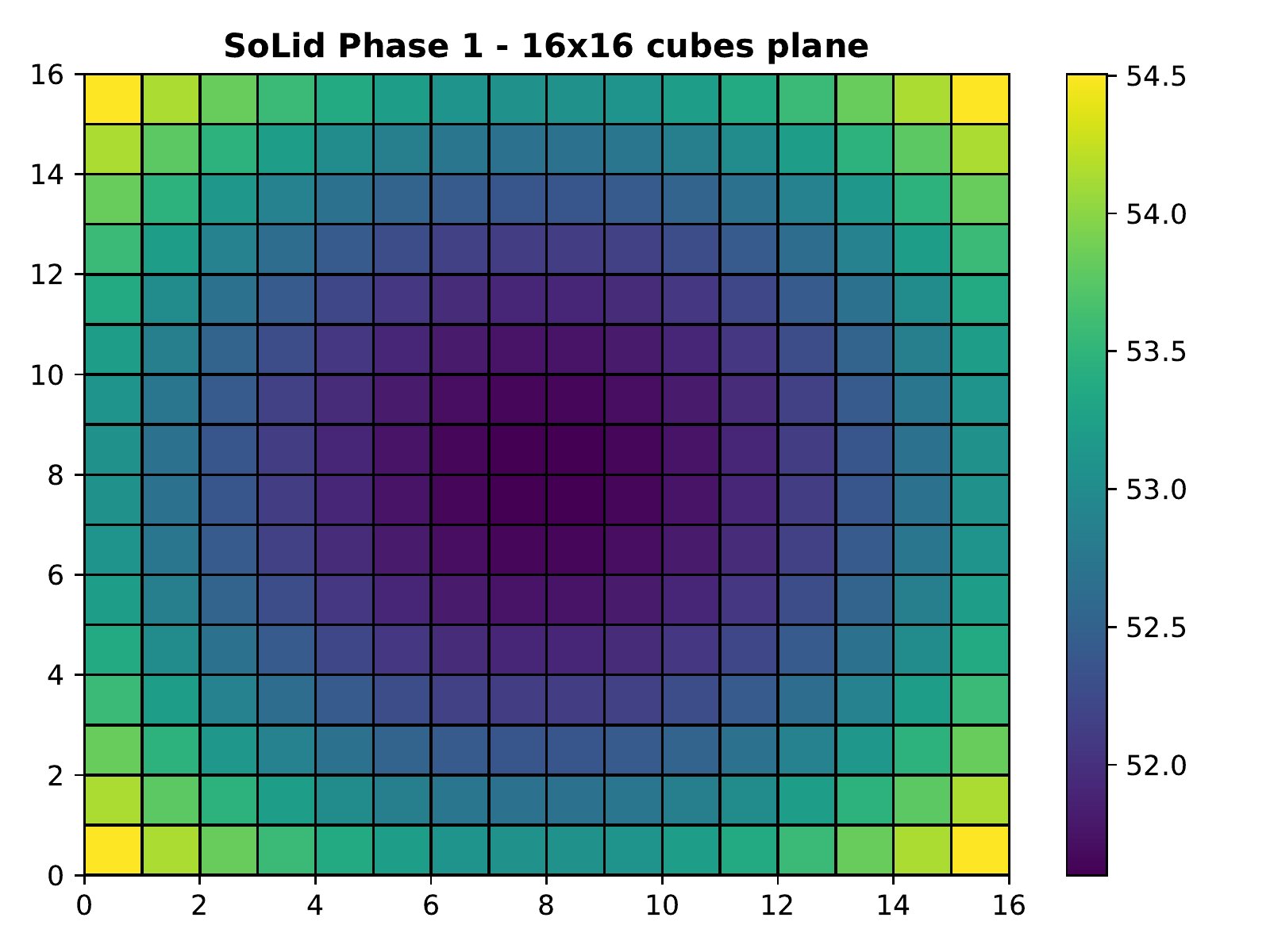}
  \caption{\label{fig:cube-ly-maps} 16$\times$16 cubes detector plane light yield maps for SM1 (left) and SoLid Phase~1 (right) extrapolated from the light yield measurements presented in this article. The average light yield is 18.9 and 52.3~PA/MeV for SM1 and SoLid Phase~1, respectively. The maximal difference, is only 6~\% for the Phase~1 compared to 43~\% for SM1.}
\end{figure}

From this last measurement for a single cube at the central position of the 16$\times$16 cubes detector plane and the attenuation length measurements (section~\ref{subsec:optical-fibres}) we can build the 2D light yield maps for the SM1 and SoLid Phase~1 16$\times$16 cubes planes. These are shown in figure~\ref{fig:cube-ly-maps}. For the SM1 configuration the average light yield of a plane is 19.0~PA/MeV, with values ranging between 16.1 to 23.1~PA/MeV. The difference between these two extreme values is 43~\%. The SM1 figure also illustrates the alternated fibre readout (up/down for vertical fibres and left/right for horizontal ones) that was decided to diffuse the non uniformities in the detector plane. For the Phase~1 configuration we observe a much more uniform light yield in the plane with only 6~\% difference between the most extreme light yields (51.6 and 54.5~PA/MeV). The average value over the plane is 52.3~PA/MeV. This illustrates the strong improvement in light yield and uniformity expected for the SoLid Phase~1 detector.

\section{Conclusion}
\label{sec:conclusion}

A precision test bench based on a \up{207}Bi calibration source developed to improve the light yield of the SoLid detector has been presented in this article. A trigger system selecting the 1~MeV conversion electrons provides a Gaussian energy peak and allows for precise comparisons of the different detector configurations that were considered. The systematic studies have shown an uncertainty on the light yield measurements of 5~\%. The light yield of the SM1 prototype has been measured to be 18.6~PA/MeV on this test bench while the observed value in the real detector was 24~PA/MeV. This 30~\% higher efficiency is attributed to the improved reflectivity for cube elements assembled in a real-scale detector module. The reactor antineutrino energy is measured through the energy deposited by the positron produced in the inverse beta decay interaction of the antineutrino in the plastic scintillator of the SoLid detector. The light yield of the Phase~1 cubes has been improved compared to the SM1 detector by a better scintillator machining (+10~\%), the choice of wrapping material (+10~\%), the type of fibre (+15~\%), the position of the \up{6}LiF:ZnS(Ag) screen (-3~\%), the number of optical fibres (+43~\%) and the type of mirror at the end of the fibre (+7~\%). The overall gain results in an expected light yield of 52$\pm$2~PA/MeV for the SoLid Phase~1 detector. This is an improvement of almost a factor 2.8, or 180~\%, in the light yield for one cube of the new detector. With this light yield the energy resolution target of $\sigma_E/E = 14$~\% at 1~MeV should be possible to achieve. The light yield uniformity of a Phase~1 detector plane, which consists of 16$\times$16 cubes, has also been improved to only 6~\% difference between the most extreme cube positions.

\section*{Acknowledgments}

This work was supported by the following funding agencies: Agence Nationale de la Recherche grant ANR-16CE31001803, Institut Carnot Mines, CNRS/IN2P3 and Region Pays de Loire in France and FWO-Vlaanderen and the Vlaamse Herculesstichting in Belgium. The United Kingdom groups acknowledge the support of the Science \& Technology Facilities Council (STFC). We are grateful for the early support given by the sub-department of Particle Physics and Merton College at Oxford and High Energy Physics at Imperial College London. We thank also our colleagues, the administrative and technical staffs of the SCK\raisebox{-0.8ex}{\scalebox{2.8}{$\cdot$}}CEN for their invaluable support for this project. Individuals have received support from the FWO-Vlaanderen and the Belgian Federal Science Policy Office (BelSpo) under the IUAP network programme. The STFC Rutherford Fellowship program and the European Research Council under the European Union's Horizon 2020 Programme (H2020-CoG)/ERC Grant Agreement n. 682474.


\end{document}

%% file: SoLid_author_list_june2018.tex
\collaboration{The SoLid Collaboration}
\author[a]{Y.~Abreu,}
\author[i]{Y.~Amhis,}
\author[a]{W.~Beaumont,}
\author[i]{M.~Bongrand,}
\author[i]{D.~Boursette,}
\author[j]{B.~C.~Castle,}
\author[b]{K.~Clark,}
\author[k]{B.~Coup\'e,}
\author[b]{D.~Cussans,}
\author[a,e]{A.~De Roeck,}
\author[d]{D.~Durand,}
\author[h]{M.~Fallot,}
\author[k]{L.~Ghys,}
\author[h]{L.~Giot,}
\author[g]{K.~Graves,}
\author[d]{B.~Guillon,}
\author[h]{D.~Henaff,}
\author[g]{B.~Hosseini,}
\author[g]{S.~Ihantola,}
\author[i]{S.~Jenzer,}
\author[k]{S.~Kalcheva,}
\author[c]{L.~N.~Kalousis,}
\author[f]{M.~Labare,}
\author[d]{G.~Lehaut,}
\author[b]{S.~Manley,}
\author[i]{L.~Manzanillas,}
\author[k]{J.~Mermans,}
\author[f]{I.~Michiels,}
\author[f,k]{C.~Moortgat,}
\author[b,m]{D.~Newbold,}
\author[l]{J.~Park,}
\author[d]{V.~Pestel,}
\author[b]{K.~Petridis,}
\author[a]{I.~Pi\~nera,}
\author[k]{L.~Popescu,}
\author[f]{D.~Ryckbosch,}
\author[j]{N.~Ryder,}
\author[g]{D.~Saunders,}
\author[i]{M.-H.~Schune,}
\author[h]{M.~Settimo,}
\author[i,n]{L.~Simard,}
\author[g]{A.~Vacheret,}
\author[f]{G.~Vandierendonck,}
\author[k]{S.~Van Dyck,}
\author[c]{P.~Van Mulders,}
\author[a]{N.~van Remortel,}
\author[a,c]{S.~Vercaemer,}
\author[a]{M.~Verstraeten,}
\author[h]{B.~Viaud,}
\author[j,m]{A.~Weber,}
\author[h]{F.~Yermia}

\affiliation[a]{Universiteit Antwerpen, Antwerpen, Belgium}
\affiliation[b]{University of Bristol, Bristol, UK}
\affiliation[c]{Vrije Universiteit Brussel, Brussel, Belgium}
\affiliation[d]{Normandie Univ, ENSICAEN, UNICAEN, CNRS/IN2P3, LPC Caen, 14000 Caen, France}
\affiliation[e]{CERN, 1211 Geneva 23, Switzerland}
\affiliation[f]{Universiteit Gent, Gent, Belgium}
\affiliation[g]{Imperial College London, Department of Physics, London, United Kingdom}
\affiliation[h]{SUBATECH, CNRS/IN2P3, Universit\'e de Nantes, Ecole des Mines de Nantes, Nantes, France}
\affiliation[i]{LAL, Univ Paris-Sud, CNRS/IN2P3, Universit\'e Paris-Saclay, Orsay, France}
\affiliation[j]{University of Oxford, Oxford, UK}
\affiliation[k]{SCK-CEN, Belgian Nuclear Research Centre, Mol, Belgium}
\affiliation[l]{Center for Neutrino Physics, Virginia Tech, Blacksburg, Virginia, 24061, USA}
\affiliation[m]{STFC, Rutherford Appleton Laboratory, Harwell Oxford, United Kingdom}
\affiliation[n]{Institut Universitaire de France, F-75005 Paris, France}